\documentclass[reprint, amsmath, amssymb, aps]{revtex4-1}
\usepackage{graphicx}
\usepackage{dcolumn}
\usepackage{bm}
\usepackage{bbold}
\usepackage{slashed}
\usepackage{soul}
\usepackage{times}
\usepackage{epstopdf}
\usepackage{epsfig}
\usepackage{graphicx}
\graphicspath{ {images/} }
\usepackage[titletoc, title]{appendix}
\usepackage[colorlinks,
            linkcolor=red,
            anchorcolor=blue,
            citecolor=green
            ]{hyperref}

\begin{document}

\title{The interplay of the sign problem and the infinite volume limit: gauge theories with a theta term}
\author{Yiming Cai}
\email{yiming@umd.edu}
\author{Thomas Cohen}
\email{cohen@physics.umd.edu}
\affiliation{Maryland Center for Fundamental Physics and the Department of Physics,\\
University of Maryland, College Park, MD, USA\\}
\author{Ari Goldbloom-Helzner}
\affiliation{Montgomery Blair High School, Silver Spring, MD, USA}
\author{Brian McPeak}
\email{bmcpeak@umich.edu}
\affiliation{Department of Physics, \\ University of Michigan,  Ann Arbor, MI, USA} 

\date{\today}

\begin{abstract}
QCD and related gauge theories  have a sign problem when a $\theta$-term is included; this complicates the extraction of physical information from Euclidean space calculations as one would do in lattice studies.   The sign problem arises in this system because the partition function for configurations with fixed topological charge $Q$, $\mathcal{Z}_Q$, are summed weighted by $\exp(i Q \theta)$ to obtain the partition function for fixed $\theta$, $ \mathcal{Z}(\theta)$.  The sign problem gets exponentially worse numerically as the space-time volume is increased.   Here it is shown that apart from the practical numerical issues associated with large volumes, there are some interesting issues of principle.   A key quantity is the energy density as a function of $\theta$, $\varepsilon(\theta) =  -\log \left( \mathcal{Z}(\theta) \right )/V$.  This is expected to be well defined in the large 4-volume limit.  Similarly, one expects the energy density for a fixed topological density $\tilde{\varepsilon}(Q/V) =  -\log \left(\mathcal{Z}_Q \right )/V$ to be well defined in the limit of large 4-volumes.  Intuitively, one might expect that  if one had the infinite volume expression for $\tilde{\varepsilon}(Q/V)$ to arbitrary accuracy, then one could reconstruct $\varepsilon(\theta)$ by directly summing over the topological sectors of the partition function.  We show here that there are circumstances where this is not the case.  In particular, this occurs in regions where the curvature of $\varepsilon(\theta)$ is negative.
\end{abstract}

\maketitle

\section{Introduction}
\subsection{$\theta$ dependence}

 Due its nonpertubative structure, the vacuum of Quantum chromodynamics (QCD) and other nonabelian gauge theories is complicated.  Accordingly, it is important to understand this vacuum structure by studying how the vacuum responds when conditions are altered.  This paper focuses on the effects of changing the so-called $\theta$ term.  The QCD Lagrangian density has the form :
 \begin{align}\label{lagrange}
  \mathcal{L}=\bar{\psi}(i\gamma^{\mu}D_{\mu}-m)\psi-\frac{1}{4}G_{\mu\nu}G^{\mu\nu} - \frac{g^2}{32\pi^2}\theta \epsilon^{\alpha\beta\mu\nu}G_{\alpha\beta}G^{\mu\nu},
  \end{align}
 where $G_{\mu \nu}$ is the field strength tensor; the last term is often omitted.  It is the so called $\theta$ term; the $\theta$ parameter is  sometime referred to as the vacuum angle.  In Euclidean space, it is  associated with a winding number Q, called topological charge, which is given by 
 \begin{align}
   Q = \int_V \ \frac{g^2}{32\pi^2} \epsilon^{\alpha \beta \mu \nu}G_{\alpha \beta}G_{\mu \nu}.
 \end{align}

 The integer $Q$ equals the difference in the number of right-handed and left-handed zero modes of the Dirac operator according to the Atiyah-Singer index theorem\cite{Indexth}. The $\theta$ term violates  CP, so the parameter $\theta$ measures the amount of CP-violation in QCD and QCD-like theories. Since $Q$ is quantized for any configuration, the $\theta$ dependence of any physical observable is periodic in $2\pi$;  thus, it is useful to restrict our attention to  $\theta$  between $-\pi$ and $\pi$.  The case of $\theta=\pi$ is particularly interesting since formally it is CP conserving---under CP transformations $\theta \rightarrow -\theta$ but $\pi$ and $-\pi$ are $2 \pi$ apart and by periodicity are equivalent.

 The $\theta$ term is of both theoretical and experimental interest because for many reasons, not the least of which because it breaks  both P and CP. On the theoretical side, the study of $\theta$ dependance is important as it gives an important indication of how the theory responds to a P and CP violating probe.  It is interesting to know, for example, whether CP is spontaneously broken in a particular theory at  $\theta=\pi$.   There are limiting cases where there are good reasons to believe that Dashen's phenomenon, the spontaneous broken of CP symmetry at $\theta=\pi$\cite{Dashen} can occur.  This is believed to happen for example in pure Yang-Mills in the large $N_c$ limit, \cite{Witten1980, witten2} or in QCD with  $N_f = 2$ degenerate light flavors with a mass small enough so that leading term in chiral perturbation theory dominates\cite{Brower}.

A critical issue for standard model physics is that, while CP violation has been observed in the electroweak sector, no CP violating effects have been observed in strong interactions. Precise measurements of the electric dipole moment of the neutron have put the upper bound of the theta term at about $10^{-9}$ away from the CP conserving point \cite{Neutron1,Neutron2,Neutron3}. The problem why $\theta$ is so tiny, so that the CP-violation is not observed, is known as the strong CP problem\cite{Dine}.  Attempts to solve the strong CP problem by invoking physics beyond the standard model remains a central problem in contemporary physics; however, it is beyond the scope of this paper.  

However, from the perspective of QCD itself, there is a related issue.  While CP violation is known to be small in QCD, one still cannot rigorously rule out the possibility that $\theta \approx \pi$ rather than  $\theta \approx 0$. The upper bound of $|\theta|<10^{-9}$  is fixed since the value of $\theta$ is proportional to CP violation, which, in turn fixes the neutron  electric dipole moment experiment.  The electric dipole moment is experimentally bounded.    However, we know $\theta=\pi$ is also formally CP invariant, so $|\theta-\pi|<10^{-9}$ may also not be in conflict with the experimental results.  This seems quite unlikely since lattice calculations done with $\theta=0$ appear to describe the world quite well.  However, in the absence of lattice studies at $\theta=\pi$, as logical matter one cannot rule out the possibility that the $\theta=\pi$ results for most observables are  close enough to  $\theta=0$ results that $\theta=\pi$ is not  excluded.

One could easily rule out $\theta=\pi$ if CP is spontaneously broken  at $\theta=\pi$\cite{Dashen}.  Thus, it would be very useful to know whether QCD (and other gauge theories) spontaneously breaks CP at  $\theta=\pi$.   If CP is spontaneously broken, then the energy density as a function of $\theta$,  $\varepsilon(\theta)$  will have a discontinuity in its slope at $\theta=\pi$, or to be more precise, it will develop such a discontinuity in the limit that the infinite volume limit is taken.      As noted above, there are regimes where spontaneous CP violation is expected to occur at $\theta=\pi$, such as a regime of infinite $N_c$ or sufficiently small quark mass with two or more flavors.     We note, however, that while both of these cases act to suggest that QCD with three colors and physical quark masses spontaneously breaks CP at $\theta=\pi$, they are by no means definitive.

For example, Witten points out that in the large $N_c$ limit, $\varepsilon (\theta)$ is parabolic as higher terms in the curvature are supressd by factors of $1/N_c$, and the periodicity condition forces the function to be defined piecewise as $\min\sum_k \, (\theta - 2 \pi k)^2$ \cite{Witten1980, witten2}.   This leads to a discontinuity at $\theta=\pi$.   However, one could imagine the following scenario: at very large but finite that the curve is very nearly a perfect parabola, except in a region of a size which goes to zero as $N_c$ goes to infinity where the curve rapidly turns over.  In such a scenario, the infinite $N_c$ theory has a discontinuity and spontaneously breaks CP while for {\it any} finite $N_c$ CP is unbroken.  Similarly,  near the chiral limit---where the first nontrivial term in the chiral expansion dominates--- it has been shown that the energy density, periodic in $2\pi$, is proportional to $(1 - \cos{\frac{\theta}{N_f}})$ in $\theta \in [-\pi, \pi]$, where $N_f$ identical to the number of degenerate light quark flavors \cite{Brower}.  This automatically yields Dashen's phenomenon, a discontinuity at $\theta=\pi$, when $N_f\geq2$.   This may seem to be compelling since, in the real world $m_q$ is small.  However, one might worry  that although small, $m_q$ may not be negligibly small.  This worry stems in part from the fact that the behavior at large $N_c$ and small $m_q$ are qualitatively different, implying that the large $N_c$ limit and the small $m_q$ limits do not commute.   This is hardly surprising, there are many places in QCD where the large $N_c$ and chiral limits do not commute\cite{Commute}.  The key point here is that the fact that an $N_c$ of three might be sufficiently large to push the system out of the small $m_q$ regime so far as the behavior at $\theta=\pi$ is concerned.

\begin{figure}[h]
    \centering
    \includegraphics[width=0.5\textwidth]{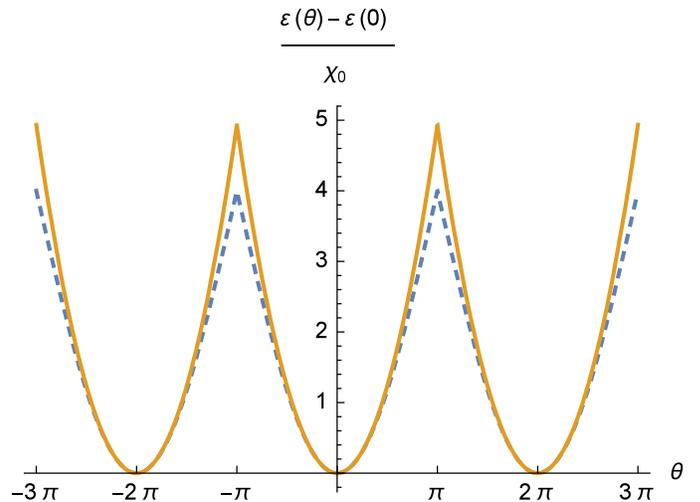}
    \caption{$\varepsilon(\theta)$  in units of the topological susceptibility and with $\varepsilon(0)$ subtracted off. Dashed line is at the large $N_c$ limit and solid line is at leading nontrivial order in a chiral expansion for two degenerate light flavors.}
    \label{fig:2limits}
 \end{figure}

To illustrate the issue, consider Fig.\ref{fig:2limits}. In this figure the form of $\varepsilon(\theta)$ is given for the infinite $N_c$ limit and for the leading nontrivial term in the chiral expansion.  To simplify the comparison (and all comparisons in this paper) we give $\varepsilon(\theta)$  divided by the topological susceptibility, $\chi_{0}=\frac{\partial^2 \varepsilon(\theta)}{\partial \theta^2}|_{\theta=0}$; we also set $\varepsilon(0)$  to zero.    It is clear that both curves have discontinuities in the slope at $\theta=\pi$ and thus both break CP spontaneously.  However, it is also clear that the two curves are quite different in the regime near $\theta=\pi$.   Thus, in the regime where $m_q$ is small and $N_c$ is simultaneously large, one expects QCD  to interpolate between these two in some manner that depends on how close the system is to the two limits.   Now, while it is clear that each of these limits has a discontinuity in the slope at $\theta=\pi$, it is {\it not} clear {\it a priori} that the interpolating function also does.  It is easy to envision a scenario in which all of the interpolating curves are smooth everywhere with no discontinuity in the slope at $\theta=\pi$ but which have a curvature at $\pi$ that increases as the limiting cases are approached and diverge at the limits yielding the sharp forms seen.

Given the importance of understanding the energy dependence of the vacuum as a function of $\theta$, it is natural to explore the prospects of calculating it directly from QCD using lattice methods.  As is well known, there is no practical way to do this due to the so-called sign problem.   This is generally considered as a practical problem associate with the natural algorithms to compute the energy density.  In this paper, we note that there is in an interesting theoretical issue connected to the sign problem.

\subsection{A sign problem}

Before outlining the key issues, it is useful to  define the quantities of interest.   We do so in Euclidean space, the natural setup for lattice studies and for simplicity of discussion work in the continuum limit  here.  We note in passing that there are subtleties associated with topology when considering the continuum limit of a discrete lattice\cite{Discrete1,Discrete2,Discrete3,Discrete4}.  However, these are unrelated to the issues discussed in this paper.  The energy density of the  QCD $\theta$-vacuum can be given in terms of the  QCD partition function in Euclidean space $\mathcal{Z}(\theta,V)$ as:
\begin{align}\label{relation1}
  \varepsilon(\theta) = -\lim_{V \rightarrow \infty} \frac{1}{V} \log{\mathcal{Z}(\theta,V)}.
 \end{align}
The Euclidean space QCD partition function is given as a functional integral
 \begin{align}
  \mathcal{Z}(\theta,V) = \int \, [dA] \, \det[ i \slashed{D}[A] - M] \, \exp(-S_{YM} + i \theta Q), \label{Integrand}
 \end{align}
 where $S_{YM}$ and $Q$ are functionals of the gluon field configurations on a space-time region of volume $V$.

 We consider the Euclidean-space theory confined to a box of four dimensional space-time volume $ V = L_x L_y L_z L_t $. Ultimately we are interested in infinite volumes. However, for practical calculations on a lattice one must use a finite $V$ and then take it to be large enough to reduce the effect of finite volume effects and estimate, with some accuracy, their size. The boundary conditions imposed on lattice are often taken to be periodic for boson fields and anti-periodic for fermions.

 It is well-known that  $\mathcal{Z}(\theta,V)$ can be written as a Fourier series over partition function $\mathcal{Z}_Q(V)$ with fixed topological charge $Q$:
 \begin{align}
  \mathcal{Z}(\theta,V) & = \sum_{Q \in \mathbb{Z}} \ \mathcal{Z}_Q(V) \ e^{i \theta Q} \nonumber\\ &= \sum_{Q \in \mathbb{Z}} \ \mathcal{Z}_Q(V) \ \cos(\theta Q),
\label{relation2}\\
    \mathcal{Z}_Q(V) &= \frac{1}{\pi}\int_0^{\pi} \mathcal{Z}(\theta) \, e^{-i \theta Q} \, \text{d} \theta \nonumber \\ & = \frac{1}{\pi}\int_0^{\pi} \mathcal{Z}(\theta) \, \cos(\theta Q) \, \text{d} \theta, \label{relation4}
 \end{align}
 where we use the fact that $\mathcal{Z}(\theta)$ is even and periodic in $\theta$ \cite{Witten1}. The fixed-topology partition function used here is \cite{Brower}
 \begin{align}\label{relation3}
     \mathcal{Z}_Q (V) = \int \, [dA]_Q \, \det[ i \slashed{D}[A] - M] \, \exp( -S_{YM} ),
 \end{align}
 where the integration is performed over all field configurations on lattice with a given topological charge Q.

 The quantity $\mathcal{Z}_Q(V) $ allows us to define a new energy density distribution function:
 \begin{align}\label{infinite}
    \tilde{\varepsilon}(q,V) = -\frac{1}{V}\log(\mathcal{Z}_{(q V)}(V) ),
 \end{align}
 where we define topological charge density $q\equiv \frac{Q}{V}$ with it understood that $q V$ is an integer, so $\varepsilon(\theta)$  and $\tilde{\varepsilon}(q,V)$ are related by:
 \begin{align}\label{thetaqlattice}
   \varepsilon(\theta)= -\lim_{V\to\infty} \frac{1}{V}\log\left(  \sum_{Q=-\infty}^\infty \ e^{-\tilde{\varepsilon}(\frac{Q}{V},V)V} \, e^{i \theta Q} \right ).
 \end{align}

One expects that the energy density should be an intensive property that depends on the intensive quantity $q=\frac{Q}{V}$.  Thus,  one expects a well-defined infinite volume limit, so that we can define:
 \begin{align}\label{relation5}
    \tilde{\varepsilon}(q) &\equiv \lim_{V\to\infty} \tilde{\varepsilon}(q,V)\; \; {\rm with}\; q V \; {\rm a \; positive\; integer}\\
     &=-\lim_{V\to\infty} \frac{1}{V}\log \left (\frac{1}{\pi}\int_0^{\pi} e^{-\varepsilon(\theta)V} \, e^{-i \theta q V} \, \text{d} \theta \right ) \, . \nonumber
 \end{align}

 Clearly, we would like to be able to compute $\varepsilon(\theta)$  directly from QCD.  If we could, we could potentially rule out definitively, the possibility the $\theta=\pi$  by showing explicitly that CP is spontaneously broken at $\theta=\pi$.  Ideally, we could answer the question of whether this happens by doing lattice studies at non-zero $\theta$.   However, lattice studies at $\theta$ away from zero are not practical and as a result $\varepsilon(\theta)$ of real QCD  remains unknown.  The reason they are not is because of a so called sign problem.

The core of the problem is the oscillatory nature of the functional integrand in Eq.~(\ref{Integrand}) or equivalently the terms in the sum in  Eq.~(\ref{relation2}). Such integrals or sums involve large cancellations which lead to the loss of a considerable amount of accuracy. This problem also makes the standard Monte Carlo methods impractical: functional integrals with an oscillatory integrand suffers from an integration weight which is not necessarily positive, and in this case normal sampling methods are not practical.  Indeed, for this problem the sign problem implies that the computation cost is exponentially large as a function of $V$.

It is worth recalling why the sign problem implies  exponentially expense in terms of computer resources.  For simplycity we consider the difference between $\varepsilon(\pi)$ and $\varepsilon(0)$. It instructive to separate the $Q=0$ term from the rest and rewrite Eq.~(\ref{thetaqlattice}) as
 \begin{align}
   \varepsilon(\pi)-  \varepsilon(0) & =   -\lim_{V\to\infty} \frac{1}{V}\log \left ( \frac{A(V)-B(V)}{A(V)}  \right ) \nonumber \\
   {\rm with} \; \; A(V) &\equiv \frac{\mathcal{Z}_{Q=0}(V)}{\mathcal{Z}(\theta=0,V)}  \nonumber\\
  {\rm and} \; \;  B(V) & \equiv -\frac{2 \sum_{Q=1}^{\infty} (-1)^Q \mathcal{Z}_{Q}(V)}{\mathcal{Z}(\theta=0,V) } \label{AB},
 \end{align}
 where the factor of 2 and sum over positive $Q$  in the definition of $B(V)$ reflects the fact that $CP$ invariance implies that $\tilde{\varepsilon}\left (\frac{-Q}{V},V \right)=\tilde{\varepsilon}\left (\frac{Q}{V},V \right)$.  In deriving Eq.~(\ref{AB}), we used the well known fact \cite{Leutwyler} that
\begin{equation}
\lim_{V \rightarrow \infty} \mathcal{Z}_{Q=0}(V)  = \lim_{V \rightarrow \infty} \mathcal{Z}(\theta=0,V) \, \label{fact} ,
\end{equation}
The Eq.~(\ref{fact}) implies that $A(V)$ is a subexponential function of $V$. This in turn implies that in order to capture the difference between $\varepsilon(\pi)$ and $\varepsilon(0)$, $B(V)$ must cancel $A(V)$ to one part in $\exp\left (V\left ( \varepsilon(\theta)-  \varepsilon(0) \right ) \right )$.   If one assumes that these differences in $\varepsilon$ are of order unity, one sees that to extract the energy dependence via a direct summation of the Fourier series, requires cancellations  due to the fluctuating  sign that scale exponentially in the volume.   This in turn, implies that to get a meaningful result one would need to compute both $A$ and $B$ with an accuracy that also scales exponentially in the volume.  But,  in a Monte Carlo algorithm the accuracy scales as the square root of the resources so to get sufficient accuracy in each Q sector and summing over sectors requires resources that scale exponentially with the volume.

This exponentially serious sign problem implies that as a practical matter, simply using Eq.(\ref{thetaqlattice}) to get $\varepsilon(\theta)$ is not practical except for very small systems that are well away from describing the infinite volume result in 3+1 dimension.  A similar exponentially serious  sign problem also occurs in QCD with a  nonzero chemical potential.
 Some possible solutions has been proposed to evade these sign problems\cite{Sasaki1,Sasaki2,Gupta,Aarts1,Aarts2,Osborn1,Osborn2,Ravagli,Sexty,Bloch}. Among all of these proposals, using an imaginary chemical potential has generated significant attention.\cite{Alford,Forcrand}. Similarly, calculating imaginary $\theta$ first and then analytically continue it to real $\theta$ in order to avoid sign problem has been used to calculate deconfinement temperature, electric dipole moment and so forth\cite{Elia,Alles1,Alles2,Horsley,Vicari,Azcoiti}. In practice, analytic continuation from imaginary $\theta$ to real $\theta$ can be done for real $\theta$ fairly near 0. This is because when $\theta$ is very small, using any reasonable expansion form of energy density, we can neglect  higher order terms in the expansion and only take the several lowest order terms as an approximate analytical form in which to extrapolate to real $\theta$. However, because we lack knowledge of the exact form for energy density $\varepsilon (\theta)$ in real QCD, it  is not practical to analytically continue imaginary $\theta$ to an arbitrary real $\theta\ \sim \pi$, where we have no reason to expect high order terms to be negligible.

The sign problem is generally thought of as a practical difficulty that prevents practical calculations of $\varepsilon(\theta)$.  However, there is also an under appreciated theoretical question associated with the sign problem that is the focus of this paper.  The issue is the following: suppose that one is able  to determine  $\tilde{\varepsilon}(q)$ with arbitrary accuracy, does this give us enough information, in principle, to reconstruct  $\varepsilon(\theta)$ by summing over topological sectors?   Intuitively, it may seem obvious at first blush that the answer is yes.  After all, both quantities are thought to be intensive and thus to be well defined in the infinite volume limit.   Thus, it seems highly plausible that while the two intensive quantities  depend  on each other,  neither should depend on finite volume corrections.    However, as will be shown here, things are a bit more subtle than this.

To pose this issue mathematically let us define the quantity
  \begin{align}
   \underline{\varepsilon}(\theta) &=  -\lim\limits_{V\to\infty} \frac{1}{V}\log\left(  \sum_{Q} \ e^{-\tilde{\varepsilon}(\frac{Q}{V})V} \, e^{i \theta Q}  \right)
   \label{limitordering}\\
   &=      -\lim\limits_{V\to\infty} \lim\limits_{\tilde V\to\infty} \frac{1}{V}\log\left (  \sum_{Q } \ e^{-\tilde{\varepsilon}(\frac{Q}{ V}, \tilde V)V} \, e^{i \theta Q} \right).\nonumber \label{limitordering} \;
 \end{align}
 The new notation $\underline{\varepsilon}(\theta)$ is used  to distinguish it with $\varepsilon(\theta)$ defined in Eq.~(\ref{thetaqlattice}).    The definitions  of $\underline{\varepsilon}(\theta)$  and $\varepsilon(\theta)$ differ only  in  an ordering of limits.  In $\underline{\varepsilon}(\theta)$, $\tilde{V}$ is taken to infinity prior to taking $V$ to infinity, while in $\varepsilon(\theta)$ the limits are taken simultaneously.  The question, then amounts to whether  or not $\underline{\varepsilon}(\theta)=\varepsilon(\theta)$.

As noted above, it seems rather plausible that $\underline{\varepsilon}(\theta)=\varepsilon(\theta)$  since it relates one intensive quantity to another. As it  turns out that in principle, one {\it can} fully reconstruct  $\varepsilon(\theta)$, from $\tilde{\varepsilon}(q)$ via directly summing over topological sectors for cases in which $\varepsilon$   curves upward ({\it i.e.}  $\varepsilon''(\theta) > 0$) everywhere in the region $-\pi < \theta  < \pi$.  Such cases are  interesting since they must spontaneously break CP at $\theta=\pi$ due to a  discontinuity in the slope (as in the examples in Fig. \ref{fig:2limits}). However, it turns out that {\it if},  $\varepsilon(\theta)$ has regions  for which   $\varepsilon''(\theta) <  0$ , then, as a result of the severe sign problem, even perfect knowledge of $\tilde{\varepsilon}(q)$ is insufficient to reconstruct those regions.  Remarkably, in these cases, in order to fully reconstruct the infinite volume behavior for $\varepsilon(\theta)$ one needs to understand the finite volume effects for  $\tilde{\varepsilon}(q,V)$.

This paper is organized as follows. In the follow section we demonstrate the phenomenon in the context of a simple ``toy'' problem, that while not being QCD, illustrates the issues.  The toy problem is one for which a dilute instanton gas is a valid approximation.   It is found that in the region where $\varepsilon''(\theta)>0$,  $\underline{\varepsilon}(\theta)=\varepsilon(\theta)$.  The key analysis in understanding this behavior is analyzed the subsequent section.  The central point is that the relevant sums can be approximated as integrals that can be analyzed in terms of a saddle point approximation.  The argument is generalized in the next section; it is shown that the phenomenon can be expected to occurs when  $\varepsilon''(\theta) \le 0$ with no restriction to the dynamics of the toy problem.    The paper concludes with a discussion of these results.

\section{A toy problem: the dilute instanton gas\label{Toy}}
 \subsection{$\varepsilon(\theta)$ versus  $\underline{\varepsilon}(\theta)$}

It is well known that neither Yang-Mills nor QCD can be approximated well by a dilute instanton gas\cite{Instanton}.  Nevertheless, we are going to first consider a dilute instanton gas with instantons of fixed action and fixed size as a toy model to illustrate the underlying issues.  We envision such a model as arising from some unspecified theory (not necessarily in 3+1 dimensions), which has instanton and anti-instanton classical solutions and an analog of the $\theta$ term and in some parametric limit the theory becomes semi-classical and is dominated by widely spaced instantons of fixed size.

In such models,  it is well known that $\mathcal{Z}_Q(V) $  is obtained by summing  over the effects of instantons and antiinstantons
\begin{equation}
\mathcal{Z}_Q(V)= \mathcal{Z}_0 \sum_{n=0}^{\infty} \frac{\left ( \frac{1}{2} \, c  \, e^{-S_0} V \right )^{|Q|+2 n}}{n! \left (n+|Q| \right)!} =\mathcal{Z}_0 I_Q (c  e^{-S_0} V),
\label{DIZQ} \end{equation}
where $\mathcal{Z}_0$ is a prefactor that sums up effects other than instantons, $S_0$ is the action of a single instanton,  $V$ is the space time volume, and $c$ is a constants with  dimension 4 that includes the effects of fluctuations.  The sum yields the modified Bessel function $I_Q (c  e^{-S_0} V)$

From this it is simple to identify $\varepsilon( \theta)$.  For integer Q
 \begin{equation}
    I_Q(z)=\frac{1}{2\pi}\int^{\pi}_{-\pi} e^{z \cos\theta} \cos(Q\theta) d\theta , \label{In}
 \end{equation}
 which from Eq. (\ref{relation2}) implies that
 \begin{equation}
  \mathcal{Z}(\theta) = \mathcal{Z}_0 \exp \left (c  e^{-S_0} V \cos(\theta) \right ),
  \end{equation}
  from which the standard dilute instanton gas expression\cite{Coleman} for  $\varepsilon(\theta)$ follows:
  \begin{align}
  \varepsilon(\theta) &= \varepsilon_0 + \chi_0 (1-\cos(\theta)) \label{standard} \\
  {\rm with} \;  \chi_0 & \equiv c  e^{-S_0} \nonumber ,
  {\rm and} \;  \varepsilon_0 \equiv \frac{- \log \left (\mathcal{Z}_0  \right )}{V} - c  e^{-S_0}  \; .\nonumber
  \end{align}

 Next let us consider $\tilde{\varepsilon}$ the energy density associated with fixed topological sectors.  Previously it was argued that this should be an intensive quantity which depends only on $q=Q/V$.  Let us verify that this is true for the dilute instanton gas model.  Start with  $ \mathcal{Z}_Q $ given in  Eq~(\ref{DIZQ})
 and exploit the series expansion  of the modified  Bessel function  $I_{\nu}(\nu z)$ around the  the uniform limit  $\nu\rightarrow\infty$ through positive real values,
\begin{widetext}
 \begin{equation}
    I_{\nu}(\nu z)\sim \frac{e^{\nu\eta}}{(2\pi\nu)^{\frac{1}{2}}(1+z^2)^{\frac{1}{4}}} \sum^{\infty}_{k=0}\frac{U_k(p)}{\nu^k} \; \; \; {\rm with} \; \;
    \eta=(1+z^2)^{\frac{1}{2}}+\log\left (\frac{z}{1+(1+z^2)^{\frac{1}{2}}} \right ), \; \;  {\rm and} \; \;
    p=(1+z^2)^{-\frac{1}{2}} ,
 \end{equation}
 where $U_k(p)$s are polynomials in $p$ of degree $3k$ with $U_0(p)$ equal to unity\cite{Bessel}.  This yields
 \begin{equation}
 \tilde{\varepsilon}(q,V) =\varepsilon_0+\chi_0+ q \log \left(\frac{q+\sqrt{\chi_0^2+q^2}}{\chi_0} \right )-\sqrt{\chi_0^2+q^2} +  \frac{\log \left ((2\pi)^2 V^2 (q^2+\chi_0^2)\right)- 4 \log \left( \sum_k\frac{U_k \left (\frac{q}{\sqrt{q^2+\chi_0^2}} \right )}{(qV)^k}\right) }{4V}.
\end{equation}
Thus, in the limit $V\rightarrow \infty$ with fixed $q$:
\begin{equation}
\tilde{\varepsilon}(q) =\lim_{V\rightarrow\infty}\tilde{\varepsilon}(q,V)=
\varepsilon_0+\chi_0+ q \log \left(\frac{q+\sqrt{\chi_0^2+q^2}}{\chi_0} \right )-\sqrt{\chi_0^2+q^2} = \varepsilon_0+\chi_0+ q \sinh^{-1} \left(\frac{q}{\chi_0} \right )-\sqrt{\chi_0^2+q^2}\; .\label{epsexpand}
\end{equation}
\end{widetext}
As advertised, $\tilde{\varepsilon}$ is an intensive quantity only dependent on $q$, a result previously derived in \cite{Leutwyler} for the analogous case of QCD with one light flavor which also has a cosinusoidal dependance of $\varepsilon(\theta)$.

Before proceeding, it is useful to note that the form of $\tilde{\varepsilon}(q)$ as a mapping of real numbers to real numbers is unique.  However, if we continue to the complex $q$ plane, as we will have cause to do later, the functional form is multi-branched with branch points at $\pm i \chi_0$ and branch cuts extending along the imaginary axis to infinity.  The imaginary axis is free of a branch cut only for $-\chi_0 < i q < \chi_0$.     The form of $\tilde{\varepsilon}(q)$ used here corresponds to  the principal branch.

 \begin{figure}[b]
    \centering
    \includegraphics[width=0.5\textwidth]{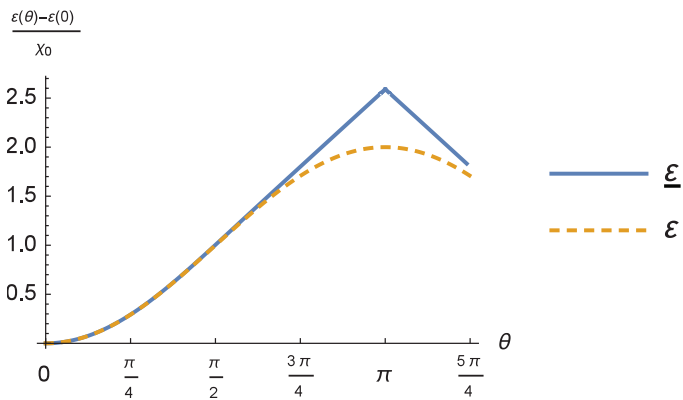}
    \caption{${\varepsilon}(\theta)$ and $ \underline{\varepsilon}(\theta)$  in units of $\chi_0$ for a dilute instanton gas model.  The two functional forms are clearly different for $\theta > \frac{\pi}{2}$ indicating that in this region knowledge of  $\tilde{\varepsilon}(q)$ is insufficient to reproduce  $ {\varepsilon}(\theta)$ by direct summation even though they are both the intensive quantities of interest. The numerical evaluation of $ \underline{\varepsilon}(\theta)$  was done with a large, but finite four dimensional volume; $V \chi_0=350$ in the calculations}
    \label{fig:cuspqlogq}
\end{figure}

With knowledge of $\tilde{\varepsilon}$, we are in a position to investigate whether it is sufficient to determine $\tilde{\varepsilon}$.  To do so we must calculate  $\underline{\varepsilon}(\theta)$, defined in Eq.~(\ref{limitordering}) and ask whether it agrees with $\varepsilon(\theta)$.  The calculation of  $\underline{\varepsilon}(\theta)$ involves evaluating $ \sum_{Q} \ e^{-\tilde{\varepsilon}(\frac{Q}{V})V} \, e^{i \theta Q}$.  Unfortunately, we know of no way to do this analytically so the calculation of $\underline{\varepsilon}(\theta)$  was done  numerically with the large $V$ limit approximated by $V$ being taken large enough so that $V \chi_0  \gg 1$ (the results were quite stable except in the immediate vicinity of $\theta=\pi/2$  by $V \chi_0=30$).   In Fig.~\ref{fig:cuspqlogq}, $\varepsilon(\theta)$ and  the numerically calculated $\underline{\varepsilon}(\theta)$ with are plotted in units of the topological susceptibility.

There are several things to notice about Fig.~\ref{fig:cuspqlogq}.  The first is that for $0<\theta < \pi/2$ one sees that up to the quality of the numerics,   $\varepsilon(\theta)=\underline{\varepsilon}(\theta)$.  However, for $\pi/2< \theta < \pi$,  $\varepsilon(\theta)\ne \underline{\varepsilon}(\theta)$.  This demonstrates quite clearly that the answer to the question of whether knowledge of $\tilde{\varepsilon} (q)$ is sufficient to reconstruct ${\varepsilon}(\theta)$ via direct summation over topological sectors---is, ``it depends.''    The focus of the reminder of this paper is on what does it depend.  Before attacking this question,  a couple of other observations are in order.  The first is that to numerical accuracy, it appears that $\underline{\varepsilon}(\theta)$ is linear in the region $\pi >\theta > \frac{\pi}{2}$.   Beyond, $\theta = \pi$ is linear with the opposite slope, yielding a discontinuous slope.

The point that separates the region for which  $\varepsilon(\theta)=\underline{\varepsilon}(\theta)$ from the region for which  $\varepsilon(\theta) \ne \underline{\varepsilon}(\theta)$ appears from the numerics to be  $\theta=\pi/2$---or something very close to it.  Assuming that the point really is exactly at $\theta=\pi/2$, gives rise to the issue of what makes that point special.  An obvious conjecture is that $\theta=\pi/2$ is a point of inflection.  As will be seen in the course of this paper, there is very strong evidence that this conjecture is correct.

\subsection{The severity of the sign problem}

One thing illustrated quite clearly by the different energy densities in Fig.~\ref{fig:cuspqlogq} for our toy problem,  the dilute instanton gas,  is just how serious the sign problem can be.  As one approaches the infinite volume limit the fractional difference between    $\tilde{\varepsilon}(q)$ and  $ \tilde{\varepsilon}(q,V) $ clearly goes to zero in the infinite volume limit.  However, even as this difference becomes vanishingly small at large volumes, their associated energy functions of $\theta$, $\underline{\varepsilon}(\theta)$  and ${\varepsilon}(\theta)$ respectively become very different for $\theta>\frac{\pi}{2}$.  Before turning to more realistic situations it is worth exploring why this is so for the toy problem.

To understand how this vanishingly small difference in the dilute instanton gas results in order unity differences between $\underline{\varepsilon}(\theta)$ and $\varepsilon(\theta)$,  it is important to first recognize that the key quantities in the calculations are not the energy densities  but the generating functions, $\mathcal{Z} \sim \exp(-V \varepsilon)$.   Note, that
the correction term of order $1/V$ in the expansion of  $ \tilde{\varepsilon}(q,V) $ in Eq.~(\ref{epsexpand}), leads to an order unity shift in $\mathcal{Z}_Q$.  One might be tempted to ascribe the order unity differences between $\underline{\varepsilon}(\theta)$ and ${\varepsilon}(\theta)$ to these differences.   However, the underlying cause is more subtle than this.

To see this, it is instructive to compute the $\theta$ dependence using various approximations to  $ \tilde{\varepsilon}(q,V) $ including various orders of correction in $1/V$ and truncating beyond it.  We define $\tilde{\varepsilon}_{n}(q,V)$ as an approximation to $ \tilde{\varepsilon}(q,V) $ that includes all terms up ${\cal O}(V^{-n})$  and truncates the rest.
Thus,
\begin{align}
 &  \tilde{\varepsilon}_{0}(q,V)= \tilde{\varepsilon}(q) \\
  &  \tilde{\varepsilon}_{1}(q,V)= \tilde{\varepsilon}(q) +  \frac{\log \left ((2\pi)^2 V^2 (q^2+\chi_0^2)\right) }{4V}   \nonumber\\
  & \tilde{\varepsilon}_{2}(q,V) =  \tilde{\varepsilon}_{1}(q,V) -  \frac{U_1 \left (\frac{q}{\sqrt{q^2+\chi_0^2}} \right )}{qV^2}  \nonumber\\
  & ... \; \; .\nonumber
 \end{align}
 This allows us to define  $\varepsilon_n(\theta)$ as
 \begin{equation}\label{epsn}
   \varepsilon_n(\theta)= -\lim_{V\to\infty} \frac{1}{V}\log\left(  \sum_{Q=-\infty}^\infty \ e^{-\tilde{\varepsilon}_n(\frac{Q}{V},V)V} \, e^{i \theta Q} \right ),
 \end{equation}
 that is, it is the standard definition expect that it replaces $\tilde{\varepsilon}(\frac{Q}{V},V)$ by $\tilde{\varepsilon}_n(\frac{Q}{V},V)$ and thereby includes some fixed level of $1/V$  corrections.   The $\varepsilon_n(\theta)$ can be computed numerically using a large but finite value for the volume.

\begin{figure}
\centering
\includegraphics[width=0.5\textwidth]{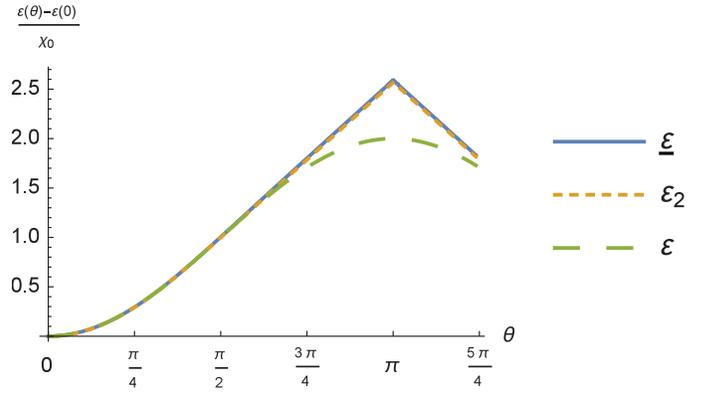}
\caption{A comparison of $\underline{\varepsilon}(\theta)$, $\varepsilon_2(\theta)$ and $\varepsilon(\theta)$. }
\label{fig:3e}
\end{figure}

 If the the explanation for the difference between $\underline{\varepsilon}(\theta)$ and ${\varepsilon}(\theta)$  were due to the order unity difference in the generating functions arising from the $1/V$  corrections in  $\tilde{\varepsilon}(q,V)$, one would find that up to numerical accuracy  $\varepsilon_n(\theta)$ would be equal to $\varepsilon(\theta)$ for all $n \ge 1$.  However, this is not the case.  In Fig.~\ref{fig:3e}, $\underline{\varepsilon}(\theta)$, ${\varepsilon_2}(\theta)$ and ${\varepsilon}(\theta)$ are plotted.  It is immediately apparent that ${\varepsilon_2}(\theta) \ne {\varepsilon}(\theta)$ differs substantially from ${\varepsilon}(\theta)$ coinciding only in  the region $0<\theta < \pi/2$; indeed it coincides with $\underline{\varepsilon}(\theta)$.  Why, then, does  $\underline{\varepsilon}(\theta)$ differ from ${\varepsilon}(\theta)$?

 Ultimately the answer resides in the discussion associated with Eq.~(\ref{AB}).  Note that to obtain ${\varepsilon}(\theta)$ accurately, one requires cancellations what are exponentially sensitive with respect to the volume.  Thus, it is not that surprising that inclusion of any given level of power law correction in the volume might be insufficient.  However, if this is the case, then why does $\underline{\varepsilon}(\theta)$ equal ${\varepsilon}(\theta)$ in the regime where  $|\theta| < \pi/2$?  The next section is aimed at answering that question in the context of this toy model.

\section{Saddle point approximation for the instanton gas model}

\subsection{A useful Identity}
To proceed, we will exploit a powerful identity relating the limit of integrals and sums which holds for a wide class of functions $f(x)$ that goes positive infinity as the  $x \rightarrow \pm \infty$ along the real axis :
\begin{equation}
\lim_{\lambda \rightarrow 0} \frac{ \int_{- \infty}^{\infty} dx \exp \left(- \lambda f\left(\frac{x}{\lambda} \right ) \right) \exp(i \theta \,  x)   }{  \sum_{n=- \infty}^{\infty} \exp \left(- \lambda f\left(\frac{n}{\lambda} \right ) \right) \exp( i \theta \,  n)  } =1, \label{identity}
\end{equation}
which holds for a large class of functions, $f$, for  $-\pi<\theta < \pi$.  The importance of this identity to our problem is that at large $V $  the sum $ \sum_{Q} \ e^{-\tilde{\varepsilon}(\frac{Q}{V})V} \, e^{i \theta Q} $ can be replaced by an integral over $Q$, which is far easier to study analytically.  The identity is not trivial.  Of course, one expects that a sum will converge to an integral in situations where the  limit of the sum yields Reimann's construction for integrals.  However, that does not happen here.  Note that the cases of interest to us here are ones in which the first term in the sum is exponentially larger than than the total even as the limit is approached.

The derivation of the identity is straightforward but is worth sketching  here as it illustrates some key features of the problem that we will exploit later.  The first step is to notice that $\sum_{n=-\infty}^\infty \delta(x-n) = \sum_{k=-\infty}^\infty \exp \left(i 2 \pi k x \right )$ from which it follows that
\begin{align}
&\sum_{n=- \infty}^{\infty} \exp \left(- \lambda f\left(\frac{n}{\lambda} \right ) \right) \exp( i \theta \,  n)  =   \label{ident1}\\
&\sum_{k=-\infty}^\infty \int_{-\infty}^ \infty dx \exp \left(- \lambda f\left(\frac{x}{\lambda} \right ) \right) \exp(i \theta x)  \exp( i  2 \pi k  x) = \nonumber \\
& \sum_{k=-\infty}^\infty  \lambda \int_{-\infty}^ \infty dy \exp \left(- \lambda \left ( f(y)  - i (\theta + 2 \pi k) \,  y \right ) \right )  ,\nonumber
\end{align}
assuming that the various integrals and  sums all converge.  The last form of Eq.~(\ref{ident1}) is instructive. Integrals  of this form in the limit of large $\lambda$ are typically  accurately approximated via the saddle point approximation assuming appropriate analyticity properties\cite{Arfken}.  Up to power law factors in $\lambda$, the $k^{\rm th}$ integral is given by
\begin{align}
& \int_{-\infty}^ \infty dy \exp \left(- \lambda \left ( f(y)  - i (\theta + 2 \pi k) \,  y \right ) \right )  \sim \rm{max}_k \exp(- \lambda g_k) \nonumber\\
& {\rm where} \; \; g_k   =\rm{min}_j (g_k^j) \; \;  {\rm with} \;  g_k^j= (f(y_k^j) - i  (\theta + 2 \pi k) \,  y_k^j ), \label{spa}
\end{align}
where $y_k^j$ is the $j^{\rm th}$ saddle point for the function $(f(y) - i  (\theta + 2 \pi k) \,  y )$ and thus satisfies:
\begin{equation}
\left .  \frac{d \left ( f(y(t)) - i ( \theta + 2 \pi k) \,  y(t) \right )}{d t }\right |_{t=t_k^j}  =0 ,
\end{equation}
where $y(t)$ is a contour in the the complex plane and $y_k^j \equiv y(t_k^j)$. Note that the derivative of both the real and imaginary parts of $f(y(t)) - i ( \theta + 2 \pi k) \,  y(t)$ have to vanish at the saddle point.  The key point is that each integral is exponentially dominated by its minimum saddle point in the complex plane as $\lambda$ becomes large.

Thus, for a wide class of functions, one  expects that at large $\lambda$, the sum over $k$ to be exponentially dominated by the term with the smallest value of $g_k$.   For many typical cases one expects this to be the $k=0$ term  for $-\pi <\theta < \pi$. In these cases, as  $\theta$ goes beyond $\pi$, one expects that the $k=-1$ term to take over as the dominant term, as $\theta$ exceeds $3 \pi$ the k=-2 to take over and so forth.  If one focuses on the  the region   $-\pi <\theta < \pi$ for these typical cases, the sum at large $\lambda$ is exponentially well approximated by the $k=0$ term.  However, from the middle line  of Eq.~(\ref{ident1}), the $k=0$ term is precisely the integral in the numerator of Eq.~(\ref{identity}) which establishes Eq.~(\ref{identity}) for this class of function.

There is an important subtlety. The saddle point approximation is obtained by distorting the path of the integral from along the real line to some other path with the same end points in the complex plane.  The integrals will coincide provided the integrand is analytic everywhere in the region enclosed by the two paths.  Distorting the path through saddle points along paths with steepest decent allows one to show that the region near the saddle  point dominates the the integral exponentially--at least locally--and thereby justify the approximation.   Saddle points are not the only points that can exponentially dominate the integrals.    Suppose the function, $f$ is not analytic everywhere, but has a branch cut singularity. In that case, the branch point can act in a manner quite analogous to a saddle point and the argument given above goes through with minor changes.  In particular, it is easy to see that circumstances can arise such that integration paths that are distorted to go around a branch point in $f$ and are arbitrarily close to the branch cut on each side can be exponentially dominated at large $\lambda$ by the contribution in the immediate vicinity of the branch point.  When this happens the branch point plays the same basic role as a saddle point in that the value of the function at the branch point can determine the value of the integral up to subexponential factors.  Note that there may be more than one branch point and the relevant branch point may be determined by the need to close contours at infinity.

Assuming that  Eq.~(\ref{identity}) holds, it can be exploited in computing  $\underline{\varepsilon}(\theta)$.   We can replace the sum over $Q$ by an integral in computing  since we are taking the infinite volume limit.  Thus,
\begin{align}
   \underline{\varepsilon}(\theta) &=  -\lim\limits_{V\to\infty} \frac{1}{V}\log\left(  \sum_{Q} \ e^{-\tilde{\varepsilon}(\frac{Q}{V})V} \, e^{i \theta Q}  \right)  \nonumber\\
  &=  -\lim\limits_{V\to\infty} \frac{1}{V}\log\left(  \int d Q \ e^{-\tilde{\varepsilon}(\frac{Q}{V})V} \, e^{i \theta Q}  \right)  \nonumber \\
   &=  -\lim\limits_{V\to\infty} \frac{1}{V}\log\left( V  \int d q \ e^{-V \left ( \tilde{\varepsilon}(q) - i \theta q  \right )} \right)  \nonumber \\
   &= \tilde{\varepsilon}(q_\theta^{\rm sp} ) - i \theta q_\theta^{\rm sp}    ,\label{sp1}
    \end{align}
where the saddle point approximation is invoked in the last equality where $q^{\rm sp}_\theta$ is the dominant saddle point associated with $\tilde{\varepsilon}(q) - i \theta q  $.

On physical grounds one expects dominant saddle points to be on the imaginary axis in $q$:  in Euclidean space, $\frac{1}{V} \, \frac{\partial \log \left(\mathcal{Z}(\theta) \right)}{\partial \theta} = i q$ which means that real $\theta$ is associated with imaginary $q$.      Note, moreover, that $\tilde{\varepsilon}(q)$ is an even function.  This means that wherever $\tilde{\varepsilon}(q)$ is analytic along the imaginary axis $\left ( \tilde{\varepsilon}(q) - i \theta q  \right )$ will be real along the imaginary axis--{\it i.e.} have a constant phase of zero.  Thus, the condition for a saddle point at $q=i q_0$  is
\begin{equation}
\left . \frac{\partial \left ( \tilde{\varepsilon}(i x) - i \theta \cdot i x \right ) }{\partial x} \right |_{x=q_0} = 0 \; .
\end{equation}
The up shot of this, plus the last  equality in Eq.~(\ref{sp1} ) is that in the infinite volume limit, one expects $\tilde{\varepsilon}$ to be related to $\underline\varepsilon$ by an analog of a Legendre transformation---but with a critical factor of $i$:
\begin{equation}\begin{split}
\tilde{\varepsilon}(q(\theta))   & =   \underline{ \varepsilon}(\theta) + i \, \theta  \, q (\theta)  \; \; \; {\rm with}  \; \; \;  q(\theta)  =  i \frac{\partial  \underline{\varepsilon}(\theta)}{\partial \theta} \\
\underline{ \varepsilon}(\theta(q))    &=   \tilde{\varepsilon}(q) -  i \, \theta(q)  \, q   \; \; \; {\rm with}  \; \; \;  \theta(q)  =  -  i \frac{\partial  \tilde{\varepsilon}(q)}{\partial q} \; .
\label{legendre} \end{split} \end{equation}

Equation (\ref{legendre}) is central to the analysis.   One key point is, that by its structure Eq.~(\ref{legendre}), requires  $\underline{ \varepsilon}(\theta)$  and $\tilde{ \varepsilon}(q)$ continued into the complex plane. It is also important to recall the limitations of Eq.~(\ref{legendre}).  Its validity requires: i)  that $\tilde{\varepsilon}(q)$ is analytic for at least some region along the imaginary axis;  ii) that the $k=0$ integral in Eq.~(\ref{ident1}) is dominated by a saddle point along the imaginary axis and  iii) that the identity in Eq.~(\ref{identity}) holds for $\tilde{\varepsilon}(q) - i \theta q $ which follows if the sum on $k$ in Eq.~(\ref{ident1}) is exponentially dominated by the $k=0$ term at large $V$.

\subsection{The saddle point approximation for the dilute instanton gas for $|\theta| < \pi/2$ \label{small theta}}

Let us return to the dilute instanton gas and focus on what the saddle point approximation tell us about the relationship of $\underline{ \varepsilon}(\theta)$ to $\varepsilon(\theta) $ for the regime $|\theta| < \pi/2$ .  We will start by assuming that the conditions justifying Eq.~(\ref{legendre}) hold and see what that implies.  Subsequently, we will argue that these conditions should hold.

 Let us test the hypothesis that in the regime $|\theta| < \pi/2$,  $\underline{ \varepsilon}(\theta) = \varepsilon(\theta)$, which for the dilute instanton approximation is given by  $\varepsilon(\theta)=\varepsilon_0 +\chi_0 \left (1-\cos(\theta) \right )$.  If this hypothesis is correct, then $\frac{\partial \underline{ \varepsilon}}{\partial \theta}= \chi_0 \sin(\theta)$,   and  the first form of Eq.~(\ref{legendre}) becomes
 \begin{equation}
 \tilde{\varepsilon}\left (i \chi_0 \sin(\theta) \right ) = \varepsilon_0 + \chi_0 \left ( 1- \cos(\theta) \right) - \chi_0 \sin(\theta) \theta .\label{consistency}
 \end{equation}
 Note that we have shown previously that for the dilute instanton gas, $\tilde{\varepsilon}(q)= \varepsilon_0+\chi_0-\sqrt{\chi_0^2+q^2} + q \sinh^{-1} \left(\frac{q}{\chi_0} \right )$.  Inserting  $q(\theta)=i \frac{\partial \underline{ \varepsilon}}{\partial \theta}=i  \chi_0 \sin(\theta)$ in this yields
 \begin{equation}
 \begin{split}
& \tilde{\varepsilon}\left (i \chi_0 \sin(\theta) \right )   \\
&=\varepsilon_0+\chi_0 (1-\sqrt{1-\sin(\theta)^2} -  \chi_0 \sin(\theta)  \sin^{-1}\left(\sin(\theta)\right )\\
& = \varepsilon_0 + \chi_0 \left ( 1- \cos(\theta) \right) - \chi_0 \sin(\theta) \theta ,\label{check}
\end{split}
 \end{equation}
where the  second equality holds for $|\theta| < \pi/2$ where $\sin^{-1} \left (\sin (\theta) \right ) = \theta$ and  $\sqrt{1-\sin(\theta)^2}=\cos(\theta)$.  Thus, in the domain $|\theta| < \pi/2$, the lefthand side of  Eq.~(\ref{consistency}) is indeed equal to the right and the hypothesis that  $\underline{ \varepsilon}(\theta) = \varepsilon(\theta)$ is consistent---providing the assumptions underlying Eq.~(\ref{legendre}) hold.

This demonstration of consistency in hardly surprising:  it justifies the empirical observation in Sec.~\ref{Toy}  based on the numerical results of direct summation with large volumes that $\underline{ \varepsilon}(\theta) = \varepsilon(\theta)$ in the regime $|\theta| < \pi/2$.  Of course, the demonstration here depends on the assumption that the three conditions justifying Eq.~(\ref{legendre}) hold.  Condition i), the existence of a region along the imaginary axis where $ \tilde{\varepsilon}(q)$  is analytic clearly holds for the form derived for the dilute instanton gas: $ \tilde{\varepsilon}(q)$ is analytic from $- i \, \chi_0 $ to $i \, \chi_0 $.  It is also highly plausible the condition ii) holds.   There is a saddle point at $q=i \chi_0 \sin(\theta)$.

 Condition iii), that the $k=0$ integral dominates the sum over $k$ in Eq.~(\ref{ident1}) requires a bit of care.  It is easy to see that there are no saddle points along the imaginary axis for any of the integrals associated with $k \ne 0$.   If saddle points exist  for $k \ne 0$,  then the second form of Eq.~(\ref{legendre}) with the substitution $\theta \rightarrow \theta + 2 \pi k$ would determine them.  Thus,
\begin{equation}
  \theta + 2 \pi k =- i  \left . \frac{\partial  \tilde  {\varepsilon}(q)}{\partial q} \right |_{q=q^{\rm sp}} ,
  \end{equation}
  Using $\tilde{\varepsilon}(q)= \varepsilon_0+\chi_0-\sqrt{\chi_0^2+q^2} + q \sinh^{-1} \left(\frac{q}{\chi_0} \right )$, this means that
  \begin{equation}
  \theta + 2 \pi k = - \sin^{-1} \left(i \, \frac{q^{\rm sp}}{\chi_0}\right )  \label{branchcond} \; .
  \end{equation}
Recall that we are considering $q^{\rm sp}$ on the imaginary axis so that  $i q^{\rm sp}$ is real and that $\tilde{\varepsilon}(q)$ corresponds to the principal branch of the function which forces the derived $ \sin^{-1} $ to also correspond to the principal branch and thus to take value from $-\frac{\pi}{2}$ to   $\frac{\pi}{2}$.  This in turn means that for $-\pi <\theta <\pi$  there are only saddle points on the imaginary axis for $k=0$  and $-\frac{\pi}{2}<\theta < \frac{\pi}{2}$.

Since physically one expects any dominant saddle point to be on the imaginary axis and there are no saddle points for $k \ne 0$ along the imaginary axis, it is highly plausible that the integrals $k \ne 0$ are dominated by regions near branch points.  Thus for these cases
\begin{equation}
\begin{split}
 {\varepsilon}_k(\theta) & \equiv -\lim\limits_{V\to\infty} \frac{1}{V}\log\left( V  \int d q \ e^{-V \left ( \tilde{\varepsilon}(q) - i (\theta + 2\pi k) q   \right )} \right)  \\
 &= \tilde{\varepsilon}(q^{\rm bp}) - (\theta + 2 \pi k ) (i q^{bp}  ) ,
 \end{split}
 \label{bp}
\end{equation}
where $q^{\rm bp}$ is the branch point and  for convenience we have introduced the symbol $\varepsilon_k(\theta) $.
For the dilute instanton gas  at $q^{\rm bp} =\pm i \chi_0$.    The relevant branch point is determined by the sign of $\theta + 2 \pi k$, since one wants to close contours at infinity.  Thus, for $k \ne 0 $ with the dilute instanton gas,
\begin{equation}
{\varepsilon}_k(\theta) = \varepsilon_0  +\chi_0+ \chi_0 ( |\theta +  2 \pi k|-\frac{\pi}{2})  \, . \label{bp2}\end{equation}
Note that in the domain $|\theta|<\frac{\pi}{2}$, ${\varepsilon}_k(\theta)$ for $k \ne 0$ is always greater than  than  $\underline{ \varepsilon}(\theta)= \varepsilon_0+ \chi_0 (1-\cos(\theta)$(which was determined (from the saddle point of the $k=0$ integral).  This means the $k \ne 0$ integrals are exponentially suppressed at large volumes compared to the $k=0$ and can be neglected, which establishes condition iii).

To summarize, for $|\theta|<\frac{\pi}{2}$ it has been shown self-consistently that at  large volumes for the dilute instanton gas, the sum over $Q$ can be replaced with an integral, this integral is dominated by a saddle point on the imaginary axis and that saddle point approximation to the integral yields $\underline{\varepsilon}(\theta) = \varepsilon(\theta) $.  This is consistent with what was observed empirically by direct numerical summation.  It also  agrees with the naive intuition that  $\varepsilon(\theta)$ ought to be directly obtainable from $\tilde{\varepsilon}(q)$ via direct summation over topological sectors given that both functions are intensive and independent of the volume.

\subsection{The saddle point approximation for the dilute instanton gas for $\pi/2<|\theta| < \pi$}

Let us consider what happens when we try to extend the analysis based on the saddle-point approximation to the $k=0$ integral for  the dilute instanton gas in the regime, $\pi/2<|\theta| < \pi$.  It should be clear that the approximation breaks down.  The easiest way to see this is is to repeat the analysis at the beginning of Subsection \ref{small theta}.  That is one can assume that $\underline{\varepsilon}(\theta)$ does equal ${\varepsilon}(\theta)$.    As in the $|\theta|<\pi/2$ case, one obtains Eq.~(\ref{consistency}) and the first equality of Eq.~(\ref{check}).  However, the second equality in Eq.~(\ref{check}) fails for $\pi/2<|\theta| < \pi$ since $\sin^{-1} \left (\sin (\theta) \right ) =\pm( \pi-\theta)$ (where $\pm$ is chosen to be the same as the sign of $\theta$) rather than $\theta$  and  $\sqrt{1-\sin(\theta)^2}=-\cos(\theta)$ rather than $\cos(\theta)$.  Thus it implies  that
\begin{equation}
\begin{split}
\tilde{ \varepsilon}\left (i \chi_0 \sin(\theta) \right ) & = \varepsilon_0 + \chi_0 \left ( 1+ \cos(\theta) \right) \pm \chi_0 \sin(\theta)(\pi- \theta) \\
& \ne \varepsilon_0 + \chi_0 \left ( 1- \cos(\theta) \right) - \chi_0 \sin(\theta)\theta ,
\end{split}\end{equation}
where the second form  is $\tilde{ \varepsilon}\left (i \chi_0 \sin(\theta) \right )$  as given in Eq.~(\ref{consistency}).  This is a clear inconsistency and implies that the starting assumption that $\underline{\varepsilon}(\theta) = {\varepsilon}(\theta)$ does not hold in this regime.This analysis shows why in the regime $\pi/2<|\theta| < \pi$, the intuitive notion that  $\underline{\varepsilon}(\theta)$ obtained from direct summation $\tilde{\varepsilon}(q)$  in the infinite volume limit should be equal ${\varepsilon}(\theta)$ fails do to an extremely severe sign problem.

There is an equivalent way to see this by working directly from  Eq.~(\ref{branchcond}) which holds for $k=0$ as well as $k \ne 0$.  By the same logic that the $k \ne 0$ integrals were shown not to have saddle points in the regime $|\theta|<\pi/2$, we can see that there are no saddle points for $k=0$ when $\pi/2<|\theta| < \pi$.  Equation (\ref{branchcond})  tell us that at saddle point  on the imaginary axis for the $k=0$ integral  for the dilute instanton gas when $ \theta  = - \sin^{-1} \left(i \, \frac{q^{\rm sp}}{\chi_0}\right )$ and the sine inverse is in its principle branch.  Since the principal branch of the inverse sign has a range from $-\pi/2$ to $\pi/2$ this can never be satisfied when $|\theta| > \pi/2$.  Thus, there are no saddle points on the imaginary axis in this regime.

In the absence of saddle points one expects the integrals to be dominated by the appropriate branch points.  The analysis of these is the same as for the $k \ne 0$ integrals in the previous subsection in the discussion accompanying Eqs~(\ref{bp}) and (\ref{bp2}).  As in that case, the branch points occur at $q^{\rm bp} = \pm i \chi_0$ and the appropriate branch point is fixed by the need to close contours at infinity.  The result in this regime is as given in Eq,~(\ref{bp2}), namely that $ \underline{\varepsilon}(\theta) = \varepsilon_0  +\chi_0+ \chi_0  |\theta |- \frac{\pi}{2} $.  This confirms that $\underline{\varepsilon}(\theta) \ne {\varepsilon}(\theta)$  in this regime and is also consistent with what was observed numerically:  $\underline{\varepsilon}$ is a linear function of $\theta$ in this  regime. It is easy to see by the same type of analysis as done in Subsection \ref{small theta}, that in this regime ${\varepsilon}_k(\theta)$ for $k \ne 0$ is always greater than  than  $\underline{ \varepsilon}(\theta)$ and thus the $k=0$ integral is exponentially dominant at large $V$.

To summarize:  for $\pi/2 <|\theta|<\pi$ it has been shown that there are no saddle points on the imaginary axis and the dominant $k=0$ integral is  dominated by a branch point.  This yields  a linear relationship:$ \underline{\varepsilon}(\theta) = \varepsilon_0  +\chi_0+ \chi_0  |\theta |-\frac{\pi}{2}  \ne     \varepsilon(\theta)$.  This result is consistent with what was observed empirically by direct numerical summation.  Moreover, it shows why the naive intuition that  $\varepsilon(\theta)$ ought to be directly obtainable from $\tilde{\varepsilon}(q)$ via direct summation over topological sectors given that both functions are intensive and independent of the volume fails.

\subsection{Analytic  Considerations}

Note that the sign problem at large volume can be viewed as an issue in analytic continuation.  In the region $|\theta| <\pi/2$ where we can reconstruct ${\varepsilon}(\theta)$ from  $\tilde{\varepsilon} (q)$ via direct summation, the calculation of ${\varepsilon}(\theta)$ amounts to analytically continuing $\tilde{\varepsilon} (q)$ to the imaginary axis and identifying the saddle points.  In this context, the sign problem can be viewed as thought of as the difficulty in doing this analytical continuation numerically  based entirely  on exponentially sensitive numerical knowledge $\tilde{\varepsilon} (q)$ on the real axis.

Issues associated with analyticity are also at the heart of an apparent puzzle.  Note that the analysis of the previous subsection and the numerical studies of the previous section  both  indicate that knowledge $\tilde{\varepsilon} (q)$---even if perfect---is insufficient to fully reconstruct  ${\varepsilon}(\theta)$ in the regime $\pi/2< \theta < \pi$ via direct summation over topological sectors for the dilute instanton gas.   Despite this, the question of whether exact knowledge of $\tilde{\varepsilon} (q)$ is sufficient to reconstruct  the full function ${\varepsilon}(\theta)$  for all $\theta$ is somewhat subtle.   Recall that the numerical evidence  and the analysis based on the saddle point approximation indicate that for  $|\theta| < \pi/2$, the knowledge of  $\tilde{\varepsilon} (q)$ {\it is} sufficient to reconstruct ${\varepsilon}(\theta)$ .  If we are able to fully reconstruct ${\varepsilon}(\theta)$ for $|\theta| < \pi/2$ from  $\tilde{\varepsilon} (q)$, one could in principle use the functional form obtained for ${\varepsilon}(\theta)$ in the regime  $|\theta| < \pi/2$ to analytically continue into the region $|\theta|>\pi/2$.   Moreover for the dilute instanton gas, the function is known to be analytic along the real axis and there is no obstruction to analytically continuing from  ${\varepsilon}(\theta)$ into the regime $\pi/2< |\theta| < \pi$.  Thus it seems as though the information contained in  $\tilde{\varepsilon} (q)$ should be sufficient to reconstruct the full function $ {\varepsilon}(\theta)$.

How can one reconcile the inability to reconstruct ${\varepsilon}(\theta)$ from $\tilde{\varepsilon} (q)$ via direct summation with the  fact  that $\tilde{\varepsilon} (q)$  contains enough information to reconstruct it?  The answer lies in the analytic structure of $\tilde{\varepsilon} (q)$ in the complex plane.  Recall that the functional form of  $\tilde{\varepsilon} (q)$  given in Eq.~(\ref{epsexpand}), while a uniquely defined function if viewed as a mapping from reals to reals, is multi-branched as a function from complex $q$ to complex $\tilde{\varepsilon}$.  Indeed, the branch points of this function played in a key role in the study of the regime $\pi/2 <|\theta|<\pi$.  The form we found  $\tilde{\varepsilon} (q)$ by studying it along the the real axis corresponds to the principal branch of this function.   However, the functional form  $\tilde{\varepsilon} (q)$  ``knows'' about all of its branches since one can analytically continue from one branch to the next.

Consider,  what happens to the function $\tilde{\varepsilon} (q)$ if one follows it from a point on the real axis along a path through the complex plane around the branch point at $q=i \chi_0$ a single time and back to the same point on the real axis evolving the function along the path via the Cauchy-Riemann equations.  One will not obtain the same value for the function--its value on the principal branch but rather the value of the function along a different branch.  More generally, as soon as the path crosses the branch cut one moves onto a new branch.  We will denote the functional form for the branch we obtain in this process as $\tilde{\tilde{\varepsilon}} (q)$; it is given by
\begin{equation}
\tilde{\tilde{\varepsilon}} (q) =  \varepsilon_0+\chi_0+ \sqrt{\chi_0^2+q^2} + q\left (i \pi - \sinh^{-1} \left(\frac{q}{\chi_0} \right )  \right) \; .\,
\end{equation}
where the inverse hyperbolic sine is taken in its principal branch;  the shift to a new branch of the inverse hyperbolic sine  is reflected by a flipped sign and the presence of the factor of $i \pi$.    It is worth noting that while $\tilde{\varepsilon} (q)$ was obtained by studying the behavior along the real $q$ axis and produced a real function, this new branch is not real for real $q$.  Thus, it cannot be obtained directly by taking the large $V$ limit of  $\tilde{\varepsilon} (q,V)$ with $q$ real (as it will be in lattice studies); it require an analytic continuation.

  It is easy to verify that if one uses this branch instead of the principal branch $\tilde{\varepsilon} (q)$ when computing the stationary phase integral for the regime $\pi/2<\theta<\pi$ then the results are indeed self-consistent and one reconstructs $\varepsilon(\theta)=\varepsilon_0 +\chi_0 \left (1-\cos(\theta) \right )$.  One can similarly verify numerically that directly summing over topological sectors using $\tilde{\tilde{\varepsilon}} (q)$  at large volume also reproduces $\varepsilon(\theta)=\varepsilon_0 +\chi_0 \left (1-\cos(\theta) \right )$ with high accuracy.
This indicates the sense in which the functional form $\tilde{\varepsilon} (q)$  contains the information to fully reconstruct ${\varepsilon}(\theta)$. It is clear that fully reconstructing it for some region of $\theta$ depends on using a particular branch of the function.  To reconstruct the full function one needs to use multiple branches---matching the appropriate branch to a given region of $\theta$.  On the other hand,  the only way we know how to extract these branches is via analytic continuation.  The bottom line, then is that while  $\tilde{\varepsilon} (q)$ contains information to fully reconstruct this information, it is in a form that is inaccessible when doing a direct sum over topological sectors.

\section{The general case}

Up to this point, we have focused on a toy model---the dilute instanton gas.  We have done so since it was solvable and hence it is possible to verify ones' analytic conclusions against direct numerical evaluations at large volume.  The real importance of the toy model  is that the analysis of it can serve as a paradigm for how to study cases which are not solvable in closed form.

The key point is that the analysis was given in the preceding section holds far more generally than for the dilute instanton gas.  It was shown there that when there exists a saddle point along the imaginary $q$ axis for $\tilde{\varepsilon}(q)- i q \theta$ then $\underline{\varepsilon}(\theta) = {\varepsilon}(\theta)$---{\it i.e.}the direct summation over topological sectors using the infinite volume extrapolation for the energy density as a function of winding number density---gives the correct infinite volume extrapolation of the energy density as a function of $\theta$.  This will hold generally.   

\begin{align}
  \underline{\varepsilon}(\theta)=  -\lim\limits_{V\to\infty} \sum_{k=-\infty}^{\infty}  \frac{1}{V}\log\left( V  \int d q \ e^{-V \left ( \tilde{\varepsilon}(q) - i (\theta+2\pi k) q  \right )} \right) , \nonumber \\
   \label{underline_e}
 \end{align}
\begin{align}
  \varepsilon(\theta)=  -\lim\limits_{V\to\infty} \sum_{k=-\infty}^{\infty}  \frac{1}{V}\log\left( V  \int d q \ e^{-V \left ( \tilde{\varepsilon}(q,V) - i (\theta+2\pi k) q  \right )} \right) . \nonumber \\
   \label{original_e}
 \end{align}

When the integral in Eq.~(\ref{underline_e}) has a saddle point $q^{sp}$ for a particular range of $\theta$, according to Eqs.~(\ref{legendre}), one should have $\underline{\varepsilon}(\theta(q^{sp}))=\tilde{\varepsilon}(q^{sp})-i\theta(q^{sp})q^{sp}$ and $\theta(q^{sp})=-i \frac{\partial \tilde{\varepsilon}(q^{sp})}{\partial q^{sp}}$. Since $\tilde{\varepsilon}(q) = \lim\limits_{V\to\infty}\tilde{\varepsilon}(q,V)$, $\theta(q^{sp})=-i \lim\limits_{V\to\infty}\frac{\partial \tilde{\varepsilon}(q^{sp},V)}{\partial q^{sp}}$ is also true, which in turns means Eq.~(\ref{original_e}) also has a saddle point at the same value of $q$ as Eq.~(\ref{underline_e}) at infinite volume limit. Then $\varepsilon(\theta(q^{sp}))=\lim\limits_{V\to\infty}\tilde{\varepsilon}(q^{sp},V)-i\theta(q^{sp})q^{sp} = \underline{\varepsilon}(\theta(q^{sp}))$, so that $\underline{\varepsilon}(\theta) = {\varepsilon}(\theta)$ is true in this general case if one assume there is a saddle point for the integral in Eq.~(\ref{underline_e}).

When there is no saddle point for integral in Eq.~(\ref{underline_e}), we know the branch point will play the similar role as the saddle point, but for Eq~(\ref{original_e}), the previous argument cannot be used here because there is no branch point.

The Legendre-like relations of Eqs.~(\ref{legendre}) will , in general, hold  if a saddle point exists regardless of the other details of the system.  This means that for a general $\theta$ dependence we can use Eqs.~(\ref{legendre}) to probe the conditions for which a saddle point exists and hence determine conditions for which $\tilde{\varepsilon}(q)$ can be used to reconstruct $\varepsilon(\theta )$ via direct summation over topological sectors.

The principal result is that for cases where  $\varepsilon(\theta)$ is analytic for real $\theta$ between $-\pi$ and $\pi$ one generically has
\begin{equation}
\underline{\varepsilon}(\theta)  =\varepsilon(\theta) \; \; {\rm if} \; \;  |\theta| <\theta_{\rm max} ,
\end{equation}
 where $\theta_{\rm max}$ is the smallest positive value of $\theta$ satisfying either of the following conditions
\begin{enumerate}
\item $\left .\frac{d^2 \varepsilon(\theta)}{d \theta^2}\right |_{\theta=\theta_{\rm max}} = 0$ . \label{cond1}
\item  $\theta_{\rm max} = \pi$ . \label{cond2}
\end{enumerate}
Moreover, if condition \ref{cond1} is satisfied, {\it i.e.} there is a point of inflection, in general $\underline{\varepsilon}(\theta)  \neq \varepsilon(\theta)$ for some region of $|\theta|> \theta_{\rm max}$ (typically extending either to $\theta=\pi$  or the next point of inflection, while if condition \ref{cond2} is satisfied then $\varepsilon(\theta)$ is not analytic at $\theta=\pi$, (typically with a discontinuous first derivative as in Fig.~\ref{fig:2limits}.)   This means that whenever there is a point of inflection in $\varepsilon(\theta)$ the sign problem is so severe that even exact knowledge of $\underline{\varepsilon}(\theta)$ is insufficient to reconstruct $\varepsilon(\theta)$ direct summation over the full range of $\theta$.

Note that this behavior is precisely what is seen for the dilute instanton gas, where $\underline{\varepsilon}(\theta)  $ equals $\varepsilon(\theta)$ before the point of inflection at $\theta=\pi/2$ and ceases to be equal beyond the point of inflection.

These conditions can be obtained by requiring self-consistency.  One starts by assuming that $\underline{\varepsilon}(\theta)  =\varepsilon(\theta)$ and then determines where this relation fails. We start by noting that   if $\underline{\varepsilon}(\theta)  =\varepsilon(\theta)$ holds for small $|\theta|$ then we expect that the the integral determining $\underline{\varepsilon}(\theta)$ is fixed by a saddle point  and not a branch point.   Were it a branch point than $\varepsilon(\theta)$  would be linear at small $\theta$ but since it is even would have to be proportional to $|\theta|$.    This is inconsistent with the hypothesis that $\varepsilon(\theta)$  is analytic from $-\pi$ to $\pi$.  Thus, one expects $\underline{\varepsilon}(\theta)  =\varepsilon(\theta)$ to continue to hold so long as the saddle point approximation remains valid and to fail when the approximation breaks down.

This breakdown occurs at a branch point beyond which the saddle point ceases to exist. The branch points are easy to identify;  $\tilde{\varepsilon}(q)$ ceases to be analytic along the imaginary axis at the branch points.

 A key identity allows to identify the branch points and establish the conditions where $\underline{\varepsilon}(\theta)  =\varepsilon(\theta)$.  The identity is that whenever  $\underline{\varepsilon}(\theta)  =\varepsilon(\theta)$ and there exists a saddle point of $\tilde{\varepsilon}(q)- i q \theta$ along the imaginary $i$ axis, then
\begin{equation}
\frac{\partial^2\varepsilon  (\theta)}{\partial \theta^2} \left .  \frac{\partial^2 \tilde{\varepsilon} (q)}{\partial q^2} \right |_{q_\theta^{\rm sp}  }  = 1 ,
\label{2ndder} \end{equation}
which can be derived directly from Eqs.~(\ref{legendre}).  When a saddle point exists the first line of  Eqs.~(\ref{legendre}) implies  that $\frac{d q^{\rm sp} }{d \theta} =  i \frac{\partial^2  {\varepsilon}(\theta)}{\partial \theta^2}$ (assuming $\underline{\varepsilon}(\theta)  =\varepsilon(\theta)$).  On the other hand, $ \frac{1}{\frac{d q^{\rm sp} }{d \theta}  }  =\frac{d \theta  }{d q^{\rm sp}}$ is fixed by the second line of  Eqs.~(\ref{legendre}): $ \frac{d \theta  }{d q^{\rm sp}}
= - i \left .\ \frac{\partial^2  \tilde {\varepsilon}(q)}{\partial q^2} \right |_{q=q^{\rm sp}}$.  Together they yield Eq.~(\ref{2ndder}).

Equation (\ref{2ndder}) implies that  $\left .\frac{\partial^2 \tilde{\varepsilon} (q)}{\partial q^2} \right |_{q_\theta^{\rm sp}  }$ diverges as one approaches a point of inflection of $\varepsilon(\theta)$ from below.  This divergence signals that $\tilde{\varepsilon} (q)$ ceases to be analytic at this point---precisely as one expects if a branch point is encountered there.  Although the branch point represents non-analytic behavior in $\tilde{\varepsilon}$ rather than $\underline{\varepsilon}$, it is clearly a point at which $\underline{\varepsilon}(\theta) $ becomes non-antalytic since the function becomes linear beyond the branch point.  This necessarily spoils the equality between  $\varepsilon(\theta)$ and  $\underline{\varepsilon}(\theta) $ since, by hypothesis  $\varepsilon(\theta)$ is analytic between $-\pi$ and $\pi$.

One might worry that, in principle,  a branch point could exist without $\left .\frac{\partial^2 \tilde{\varepsilon} (q)}{\partial q^2} \right |_{q_\theta^{\rm sp}  }$ diverging as the branch point is approached below. If this happened, one could have a breakdown of the condition $\underline{\varepsilon}(\theta)  =\varepsilon(\theta)$ without an inflection point in $\varepsilon(\theta)$.  The concern stems from the possibility that higher derivative could diverge signal the nonanalyticity even if the second derivative remains finite.   One would expect such behavior,  if for example $\tilde{\varepsilon} (q)$ had a contribution proportional to $\chi_0^{3/2}  \left( 1 + \frac{q^2}{\chi_0^2} \right )^{5/2}$; the second derivative remains finite as $q \rightarrow i  \chi_0$ but the third derivative diverges as do all higher derivatives.  However, we can rule out this possibility if $\varepsilon(\theta)$ is analytic between $-\pi$ and $\pi$.

 It is easy to show from Eqs. (\ref{2ndder}) that  $\left .\frac{\partial^n \tilde{\varepsilon} (q)}{\partial q^n} \right |_{q_\theta^{\rm sp}  }$  for $n>2$ is given by
\begin{widetext}
\begin{equation}
\left .\frac{\partial^n \tilde{\varepsilon} (q)}{\partial q^n} \right |_{q_\theta^{\rm sp}  } = \sum_{k_2,k_3,\cdots k_n}  \, \delta_{n, (2+k_2+k_3+\cdots k_n)} \, \frac{ c_{k_2,\cdots, k_n}\prod_{j=2,n} \left ( \frac{\partial^j \varepsilon  (\theta)}{\partial \theta^j}  \right )^{k_j}}{\left (  \frac{\partial^2\varepsilon  (\theta)}{\partial \theta^2}  \right )^{2n-3}}  ,\,
\label{form} \end{equation}
\end{widetext}
where $c_{k_1,k_2,\cdots, k_n}$ are calculable coefficients and the $k_j$  nonnegative integers.  The structure in Eq.~(\ref{form}) is significant since, by hypothesis the system is in a regime in which $\varepsilon  (\theta)$
is analytic and thus $\frac{\partial^j \varepsilon  (\theta)}{\partial \theta^j} $ is finite for all $j$.  Therefore, the only way that that $\left .\frac{\partial^n \tilde{\varepsilon} (q)}{\partial q^n} \right |_{q_\theta^{\rm sp}  } $ can diverge is if $\frac{\partial^2 \varepsilon(\theta)}{\partial \theta^2} =0$---{\it i.e} the system is at a point of inflection.

The conclusion of this analysis is that one expects that when $\varepsilon  (\theta)$ has a point of inflection somewhere in the domain $-\pi <\theta < \pi$,  for part of the domain the sign problem is so severe that one cannot obtain  $\varepsilon  (\theta)$ by directly sum over topological sectors using $ \tilde{\varepsilon} (q)$, the finite volume limit of  $ \tilde{\varepsilon} (q,V)$.  We tested this conclusion numerically by constructing numerous hypothetical forms  $\varepsilon  (\theta)$ that contained points of inflection for which we could obtain $ \tilde{\varepsilon} (q)$.  Using these extracted $ \tilde{\varepsilon} (q)$ we found in all cases that there were regions of $\theta$ beyond the point of inflection for which we could not reconstruct  $\varepsilon  (\theta)$ by summing over topological sectors.

\section{Conclusion}

This paper explored a subtlety in the relationship between the infinite volume limit and the sign problem in the context of theories with a $\theta$ term.  It was shown that there exist circumstances for which the sign problem is so severe that for some values of $\theta$ one can cannot obtain the  correct infinite volume $\varepsilon  (\theta)$ by summing over topological sectors using the exact infinite volume form for $ \tilde{\varepsilon} (q)$.  This occurs when  $\varepsilon  (\theta)$  has a point of inflection between $-\pi$ and $\pi$.  This can be taken as an illustration of just how serious sign problems can be.

However, it is also worth stressing that there are regions in $\theta$ where $ \tilde{\varepsilon} (q)$ is sufficient to obtain  $\varepsilon  (\theta)$ by direct summation.  In some ways this is quite remarkable.   After all, there are power law differences in the volume between $ \tilde{\varepsilon} (q,V)$, $ \tilde{\varepsilon} (q)$ which translates into order one errors in $\mathcal{Z}_Q$, while one requires the sum over $Q$ to yield cancellations which are accurate up to exponential accuracy in $V$ since individual terms $\mathcal{Z}_Q$ is exponentially larger than the sum $\mathcal{Z}(\theta)$.  Thus, one requires some type of conspiracy for these order one errors not to spoil the cancellations.  Ultimately, the reason why $ \tilde{\varepsilon} (q)$ turns out to be sufficient in these cases is that the sum over topological sectors can be rewritten as a sum of integrals, which at large $V$ are exponentially dominated by a single integral which in turn may be well approximated using the saddle point method.

One interesting mathematical fact that emerges from the analysis concerns the analytic structure of  the infinite volume functions $\varepsilon(\theta) $  , $ \tilde{\varepsilon} (q)$: unless both functions are trivially constant,  at least one of these is not analytic over the entire complex plane.  Recall that $\theta$ and $q$ have something of a conjugate relationship with each other up to  a key factor of $i$.    We found that in cases where  $\varepsilon(\theta) $ had a point of inflection that $ \tilde{\varepsilon} (q)$  had place where it was not analytic.   The branch point in $ \tilde{\varepsilon} (q)$ was approached as the point of inflection was approached in  $\varepsilon(\theta) $.  Thus, the only way that $ \tilde{\varepsilon} (q)$ can be analytic everywhere from $\-pi$ to $\pi$ is for $\varepsilon(\theta) $ to have no inflection points in the entire region; {\it i.e.} to have positive curvature.  However, if this is the case, then periodicity in $\theta$ implies discontinuity of the slope at $\theta= n \pi$, which is clearly non-analytic.

While this paper has focused on a theoretical issue, the analysis may prove of  value in extracting information about $\varepsilon(\theta)$ from practical lattice calculations.  Note that given the sign problem,  current techniques are not sufficient to directly compute  $\varepsilon(\theta)$ .  However, one might hope that the determination of whether or not  $\varepsilon(\theta)$ is in some qualitative class of functions might be accessible via a study of $ \tilde{\varepsilon} (q)$ or equivalently ${\varepsilon} (i \theta)$.  This possibility will be explored in future work.

\section{Acknowledgments}

We acknowledge the support of the U.S. Department of Energy.

\newpage

\bibliography{references}

\begin{thebibliography}{38}%
\makeatletter
\providecommand \@ifxundefined [1]{%
 \@ifx{#1\undefined}
}%
\providecommand \@ifnum [1]{%
 \ifnum #1\expandafter \@firstoftwo
 \else \expandafter \@secondoftwo
 \fi
}%
\providecommand \@ifx [1]{%
 \ifx #1\expandafter \@firstoftwo
 \else \expandafter \@secondoftwo
 \fi
}%
\providecommand \natexlab [1]{#1}%
\providecommand \enquote  [1]{``#1''}%
\providecommand \bibnamefont  [1]{#1}%
\providecommand \bibfnamefont [1]{#1}%
\providecommand \citenamefont [1]{#1}%
\providecommand \href@noop [0]{\@secondoftwo}%
\providecommand \href [0]{\begingroup \@sanitize@url \@href}%
\providecommand \@href[1]{\@@startlink{#1}\@@href}%
\providecommand \@@href[1]{\endgroup#1\@@endlink}%
\providecommand \@sanitize@url [0]{\catcode `\\12\catcode `\$12\catcode
  `\&12\catcode `\#12\catcode `\^12\catcode `\_12\catcode `\%12\relax}%
\providecommand \@@startlink[1]{}%
\providecommand \@@endlink[0]{}%
\providecommand \url  [0]{\begingroup\@sanitize@url \@url }%
\providecommand \@url [1]{\endgroup\@href {#1}{\urlprefix }}%
\providecommand \urlprefix  [0]{URL }%
\providecommand \Eprint [0]{\href }%
\providecommand \doibase [0]{http://dx.doi.org/}%
\providecommand \selectlanguage [0]{\@gobble}%
\providecommand \bibinfo  [0]{\@secondoftwo}%
\providecommand \bibfield  [0]{\@secondoftwo}%
\providecommand \translation [1]{[#1]}%
\providecommand \BibitemOpen [0]{}%
\providecommand \bibitemStop [0]{}%
\providecommand \bibitemNoStop [0]{.\EOS\space}%
\providecommand \EOS [0]{\spacefactor3000\relax}%
\providecommand \BibitemShut  [1]{\csname bibitem#1\endcsname}%
\let\auto@bib@innerbib\@empty
\bibitem [{\citenamefont {Atiyah}\ and\ \citenamefont
  {Singer}(1969)}]{Indexth}%
  \BibitemOpen
  \bibfield  {author} {\bibinfo {author} {\bibfnamefont {M.~F.}\ \bibnamefont
  {Atiyah}}\ and\ \bibinfo {author} {\bibfnamefont {I.~M.}\ \bibnamefont
  {Singer}},\ }\href {\doibase 10.1090/S0002-9904-1963-10957-X} {\bibfield
  {journal} {\bibinfo  {journal} {Bull. Am. Math. Soc.}\ }\textbf {\bibinfo
  {volume} {69}},\ \bibinfo {pages} {422} (\bibinfo {year} {1969})}\BibitemShut
  {NoStop}%
\bibitem [{\citenamefont {Dashen}(1971)}]{Dashen}%
  \BibitemOpen
  \bibfield  {author} {\bibinfo {author} {\bibfnamefont {R.~F.}\ \bibnamefont
  {Dashen}},\ }\href {\doibase 10.1103/PhysRevD.3.1879} {\bibfield  {journal}
  {\bibinfo  {journal} {Phys. Rev.}\ }\textbf {\bibinfo {volume} {D3}},\
  \bibinfo {pages} {1879} (\bibinfo {year} {1971})}\BibitemShut {NoStop}%
\bibitem [{\citenamefont {Witten}(1980)}]{Witten1980}%
  \BibitemOpen
  \bibfield  {author} {\bibinfo {author} {\bibfnamefont {E.}~\bibnamefont
  {Witten}},\ }\href {\doibase 10.1016/0003-4916(80)90325-5} {\bibfield
  {journal} {\bibinfo  {journal} {Annals Phys.}\ }\textbf {\bibinfo {volume}
  {128}},\ \bibinfo {pages} {363} (\bibinfo {year} {1980})}\BibitemShut
  {NoStop}%
\bibitem [{\citenamefont {Witten}(1979)}]{witten2}%
  \BibitemOpen
  \bibfield  {author} {\bibinfo {author} {\bibfnamefont {E.}~\bibnamefont
  {Witten}},\ }\href {\doibase 10.1016/0550-3213(79)90232-3} {\bibfield
  {journal} {\bibinfo  {journal} {Nucl. Phys.}\ }\textbf {\bibinfo {volume}
  {B160}},\ \bibinfo {pages} {57} (\bibinfo {year} {1979})}\BibitemShut
  {NoStop}%
\bibitem [{\citenamefont {Brower}\ \emph {et~al.}(2003)\citenamefont {Brower},
  \citenamefont {Chandrasekharan}, \citenamefont {Negele},\ and\ \citenamefont
  {Wiese}}]{Brower}%
  \BibitemOpen
  \bibfield  {author} {\bibinfo {author} {\bibfnamefont {R.}~\bibnamefont
  {Brower}}, \bibinfo {author} {\bibfnamefont {S.}~\bibnamefont
  {Chandrasekharan}}, \bibinfo {author} {\bibfnamefont {J.~W.}\ \bibnamefont
  {Negele}}, \ and\ \bibinfo {author} {\bibfnamefont {U.~J.}\ \bibnamefont
  {Wiese}},\ }\href {\doibase 10.1016/S0370-2693(03)00369-1} {\bibfield
  {journal} {\bibinfo  {journal} {Phys. Lett.}\ }\textbf {\bibinfo {volume}
  {B560}},\ \bibinfo {pages} {64} (\bibinfo {year} {2003})},\ \Eprint
  {http://arxiv.org/abs/hep-lat/0302005} {arXiv:hep-lat/0302005 [hep-lat]}
  \BibitemShut {NoStop}%
\bibitem [{\citenamefont {Smith}\ \emph {et~al.}(1990)\citenamefont {Smith}
  \emph {et~al.}}]{Neutron1}%
  \BibitemOpen
  \bibfield  {author} {\bibinfo {author} {\bibfnamefont {K.~F.}\ \bibnamefont
  {Smith}} \emph {et~al.},\ }\href {\doibase 10.1016/0370-2693(90)92027-G}
  {\bibfield  {journal} {\bibinfo  {journal} {Phys. Lett.}\ }\textbf {\bibinfo
  {volume} {B234}},\ \bibinfo {pages} {191} (\bibinfo {year}
  {1990})}\BibitemShut {NoStop}%
\bibitem [{\citenamefont {Altarev}\ \emph {et~al.}(1992)\citenamefont {Altarev}
  \emph {et~al.}}]{Neutron2}%
  \BibitemOpen
  \bibfield  {author} {\bibinfo {author} {\bibfnamefont {I.~S.}\ \bibnamefont
  {Altarev}} \emph {et~al.},\ }\href {\doibase 10.1016/0370-2693(92)90571-K}
  {\bibfield  {journal} {\bibinfo  {journal} {Phys. Lett.}\ }\textbf {\bibinfo
  {volume} {B276}},\ \bibinfo {pages} {242} (\bibinfo {year}
  {1992})}\BibitemShut {NoStop}%
\bibitem [{\citenamefont {Jacobs}\ \emph {et~al.}(1993)\citenamefont {Jacobs},
  \citenamefont {Klipstein}, \citenamefont {Lamoreaux}, \citenamefont
  {Heckel},\ and\ \citenamefont {Fortson}}]{Neutron3}%
  \BibitemOpen
  \bibfield  {author} {\bibinfo {author} {\bibfnamefont {J.~P.}\ \bibnamefont
  {Jacobs}}, \bibinfo {author} {\bibfnamefont {W.~M.}\ \bibnamefont
  {Klipstein}}, \bibinfo {author} {\bibfnamefont {S.~K.}\ \bibnamefont
  {Lamoreaux}}, \bibinfo {author} {\bibfnamefont {B.~R.}\ \bibnamefont
  {Heckel}}, \ and\ \bibinfo {author} {\bibfnamefont {E.~N.}\ \bibnamefont
  {Fortson}},\ }\href {\doibase 10.1103/PhysRevLett.71.3782} {\bibfield
  {journal} {\bibinfo  {journal} {Phys. Rev. Lett.}\ }\textbf {\bibinfo
  {volume} {71}},\ \bibinfo {pages} {3782} (\bibinfo {year}
  {1993})}\BibitemShut {NoStop}%
\bibitem [{\citenamefont {Dine}(2001)}]{Dine}%
  \BibitemOpen
  \bibfield  {author} {\bibinfo {author} {\bibfnamefont {M.}~\bibnamefont
  {Dine}},\ }\href@noop {} {\bibfield  {journal} {\bibinfo  {journal} {Flavor
  Physics for the Millennium: TASI 2000: Boulder, Colorado, US, 4-30 June
  2000}\ ,\ \bibinfo {pages} {349}} (\bibinfo {year} {2001})}\BibitemShut
  {NoStop}%
\bibitem [{\citenamefont {Cohen}(1996)}]{Commute}%
  \BibitemOpen
  \bibfield  {author} {\bibinfo {author} {\bibfnamefont {T.~D.}\ \bibnamefont
  {Cohen}},\ }\href {\doibase 10.1103/RevModPhys.68.599} {\bibfield  {journal}
  {\bibinfo  {journal} {Rev. Mod. Phys.}\ }\textbf {\bibinfo {volume} {68}},\
  \bibinfo {pages} {599} (\bibinfo {year} {1996})}\BibitemShut {NoStop}%
\bibitem [{\citenamefont {Smith}\ and\ \citenamefont {Teper}()}]{Discrete1}%
  \BibitemOpen
  \bibfield  {author} {\bibinfo {author} {\bibfnamefont {D.~A.}\ \bibnamefont
  {Smith}}\ and\ \bibinfo {author} {\bibfnamefont {M.~J.}\ \bibnamefont
  {Teper}},\ }\href@noop {} {\ }\BibitemShut {NoStop}%
\bibitem [{\citenamefont {Gockeler}\ \emph {et~al.}(1986)\citenamefont
  {Gockeler}, \citenamefont {Laursen}, \citenamefont {Schierholz},\ and\
  \citenamefont {Wiese}}]{Discrete2}%
  \BibitemOpen
  \bibfield  {author} {\bibinfo {author} {\bibfnamefont {M.}~\bibnamefont
  {Gockeler}}, \bibinfo {author} {\bibfnamefont {M.~L.}\ \bibnamefont
  {Laursen}}, \bibinfo {author} {\bibfnamefont {G.}~\bibnamefont {Schierholz}},
  \ and\ \bibinfo {author} {\bibfnamefont {U.~J.}\ \bibnamefont {Wiese}},\
  }\href {\doibase 10.1007/BF01221000} {\bibfield  {journal} {\bibinfo
  {journal} {Commun. Math. Phys.}\ }\textbf {\bibinfo {volume} {107}},\
  \bibinfo {pages} {467} (\bibinfo {year} {1986})}\BibitemShut {NoStop}%
\bibitem [{\citenamefont {Luscher}(1982)}]{Discrete3}%
  \BibitemOpen
  \bibfield  {author} {\bibinfo {author} {\bibfnamefont {M.}~\bibnamefont
  {Luscher}},\ }\href {\doibase 10.1007/BF02029132} {\bibfield  {journal}
  {\bibinfo  {journal} {Commun. Math. Phys.}\ }\textbf {\bibinfo {volume}
  {85}},\ \bibinfo {pages} {39} (\bibinfo {year} {1982})}\BibitemShut {NoStop}%
\bibitem [{\citenamefont {Woit}(1983)}]{Discrete4}%
  \BibitemOpen
  \bibfield  {author} {\bibinfo {author} {\bibfnamefont {P.}~\bibnamefont
  {Woit}},\ }\href {\doibase 10.1103/PhysRevLett.51.638} {\bibfield  {journal}
  {\bibinfo  {journal} {Phys. Rev. Lett.}\ }\textbf {\bibinfo {volume} {51}},\
  \bibinfo {pages} {638} (\bibinfo {year} {1983})}\BibitemShut {NoStop}%
\bibitem [{\citenamefont {Witten}(1998)}]{Witten1}%
  \BibitemOpen
  \bibfield  {author} {\bibinfo {author} {\bibfnamefont {E.}~\bibnamefont
  {Witten}},\ }\href {\doibase 10.1103/PhysRevLett.81.2862} {\bibfield
  {journal} {\bibinfo  {journal} {Phys. Rev. Lett.}\ }\textbf {\bibinfo
  {volume} {81}},\ \bibinfo {pages} {2862} (\bibinfo {year} {1998})},\ \Eprint
  {http://arxiv.org/abs/hep-th/9807109} {arXiv:hep-th/9807109 [hep-th]}
  \BibitemShut {NoStop}%
\bibitem [{\citenamefont {Leutwyler}\ and\ \citenamefont
  {Smilga}(1992)}]{Leutwyler}%
  \BibitemOpen
  \bibfield  {author} {\bibinfo {author} {\bibfnamefont {H.}~\bibnamefont
  {Leutwyler}}\ and\ \bibinfo {author} {\bibfnamefont {A.~V.}\ \bibnamefont
  {Smilga}},\ }\href {\doibase 10.1103/PhysRevD.46.5607} {\bibfield  {journal}
  {\bibinfo  {journal} {Phys. Rev.}\ }\textbf {\bibinfo {volume} {D46}},\
  \bibinfo {pages} {5607} (\bibinfo {year} {1992})}\BibitemShut {NoStop}%
\bibitem [{\citenamefont {Sasaki}\ \emph {et~al.}(2012)\citenamefont {Sasaki},
  \citenamefont {Takahashi}, \citenamefont {Sakai}, \citenamefont {Kouno},\
  and\ \citenamefont {Yahiro}}]{Sasaki1}%
  \BibitemOpen
  \bibfield  {author} {\bibinfo {author} {\bibfnamefont {T.}~\bibnamefont
  {Sasaki}}, \bibinfo {author} {\bibfnamefont {J.}~\bibnamefont {Takahashi}},
  \bibinfo {author} {\bibfnamefont {Y.}~\bibnamefont {Sakai}}, \bibinfo
  {author} {\bibfnamefont {H.}~\bibnamefont {Kouno}}, \ and\ \bibinfo {author}
  {\bibfnamefont {M.}~\bibnamefont {Yahiro}},\ }\href {\doibase
  10.1103/PhysRevD.85.056009} {\bibfield  {journal} {\bibinfo  {journal} {Phys.
  Rev.}\ }\textbf {\bibinfo {volume} {D85}},\ \bibinfo {pages} {056009}
  (\bibinfo {year} {2012})},\ \Eprint {http://arxiv.org/abs/1112.6086}
  {arXiv:1112.6086 [hep-ph]} \BibitemShut {NoStop}%
\bibitem [{\citenamefont {Sasaki}\ \emph {et~al.}(2013)\citenamefont {Sasaki},
  \citenamefont {Kouno},\ and\ \citenamefont {Yahiro}}]{Sasaki2}%
  \BibitemOpen
  \bibfield  {author} {\bibinfo {author} {\bibfnamefont {T.}~\bibnamefont
  {Sasaki}}, \bibinfo {author} {\bibfnamefont {H.}~\bibnamefont {Kouno}}, \
  and\ \bibinfo {author} {\bibfnamefont {M.}~\bibnamefont {Yahiro}},\ }\href
  {\doibase 10.1103/PhysRevD.87.056003} {\bibfield  {journal} {\bibinfo
  {journal} {Phys. Rev.}\ }\textbf {\bibinfo {volume} {D87}},\ \bibinfo {pages}
  {056003} (\bibinfo {year} {2013})},\ \Eprint {http://arxiv.org/abs/1303.5169}
  {arXiv:1303.5169 [hep-ph]} \BibitemShut {NoStop}%
\bibitem [{\citenamefont {Gupta}(2004)}]{Gupta}%
  \BibitemOpen
  \bibfield  {author} {\bibinfo {author} {\bibfnamefont {S.}~\bibnamefont
  {Gupta}},\ }\href {\doibase 10.1007/BF02704891} {\bibfield  {journal}
  {\bibinfo  {journal} {Pramana}\ }\textbf {\bibinfo {volume} {63}},\ \bibinfo
  {pages} {1211} (\bibinfo {year} {2004})}\BibitemShut {NoStop}%
\bibitem [{\citenamefont {Aarts}(2009)}]{Aarts1}%
  \BibitemOpen
  \bibfield  {author} {\bibinfo {author} {\bibfnamefont {G.}~\bibnamefont
  {Aarts}},\ }\href {\doibase 10.1103/PhysRevLett.102.131601} {\bibfield
  {journal} {\bibinfo  {journal} {Phys. Rev. Lett.}\ }\textbf {\bibinfo
  {volume} {102}},\ \bibinfo {pages} {131601} (\bibinfo {year} {2009})},\
  \Eprint {http://arxiv.org/abs/0810.2089} {arXiv:0810.2089 [hep-lat]}
  \BibitemShut {NoStop}%
\bibitem [{\citenamefont {Aarts}\ and\ \citenamefont
  {Splittorff}(2010)}]{Aarts2}%
  \BibitemOpen
  \bibfield  {author} {\bibinfo {author} {\bibfnamefont {G.}~\bibnamefont
  {Aarts}}\ and\ \bibinfo {author} {\bibfnamefont {K.}~\bibnamefont
  {Splittorff}},\ }\href {\doibase 10.1007/JHEP08(2010)017} {\bibfield
  {journal} {\bibinfo  {journal} {JHEP}\ }\textbf {\bibinfo {volume} {08}},\
  \bibinfo {pages} {017} (\bibinfo {year} {2010})},\ \Eprint
  {http://arxiv.org/abs/1006.0332} {arXiv:1006.0332 [hep-lat]} \BibitemShut
  {NoStop}%
\bibitem [{\citenamefont {Osborn}\ \emph
  {et~al.}(2008{\natexlab{a}})\citenamefont {Osborn}, \citenamefont
  {Splittorff},\ and\ \citenamefont {Verbaarschot}}]{Osborn1}%
  \BibitemOpen
  \bibfield  {author} {\bibinfo {author} {\bibfnamefont {J.~C.}\ \bibnamefont
  {Osborn}}, \bibinfo {author} {\bibfnamefont {K.}~\bibnamefont {Splittorff}},
  \ and\ \bibinfo {author} {\bibfnamefont {J.~J.~M.}\ \bibnamefont
  {Verbaarschot}},\ }\href {\doibase 10.1103/PhysRevD.78.105006} {\bibfield
  {journal} {\bibinfo  {journal} {Phys. Rev.}\ }\textbf {\bibinfo {volume}
  {D78}},\ \bibinfo {pages} {105006} (\bibinfo {year} {2008}{\natexlab{a}})},\
  \Eprint {http://arxiv.org/abs/0807.4584} {arXiv:0807.4584 [hep-lat]}
  \BibitemShut {NoStop}%
\bibitem [{\citenamefont {Osborn}\ \emph
  {et~al.}(2008{\natexlab{b}})\citenamefont {Osborn}, \citenamefont
  {Splittorff},\ and\ \citenamefont {Verbaarschot}}]{Osborn2}%
  \BibitemOpen
  \bibfield  {author} {\bibinfo {author} {\bibfnamefont {J.~C.}\ \bibnamefont
  {Osborn}}, \bibinfo {author} {\bibfnamefont {K.}~\bibnamefont {Splittorff}},
  \ and\ \bibinfo {author} {\bibfnamefont {J.~J.~M.}\ \bibnamefont
  {Verbaarschot}},\ }in\ \href
  {http://inspirehep.net/record/793109/files/arXiv:0808.1982.pdf} {\emph
  {\bibinfo {booktitle} {{Continous advances in QCD. Proceedings, 8th Workshop,
  CAQCD-08, Minneapolis, USA, May 15-18, 2008}}}}\ (\bibinfo {year} {2008})\
  pp.\ \bibinfo {pages} {135--147},\ \Eprint {http://arxiv.org/abs/0808.1982}
  {arXiv:0808.1982 [hep-lat]} \BibitemShut {NoStop}%
\bibitem [{\citenamefont {Ravagli}\ and\ \citenamefont
  {Verbaarschot}(2007)}]{Ravagli}%
  \BibitemOpen
  \bibfield  {author} {\bibinfo {author} {\bibfnamefont {L.}~\bibnamefont
  {Ravagli}}\ and\ \bibinfo {author} {\bibfnamefont {J.~J.~M.}\ \bibnamefont
  {Verbaarschot}},\ }\href {\doibase 10.1103/PhysRevD.76.054506} {\bibfield
  {journal} {\bibinfo  {journal} {Phys. Rev.}\ }\textbf {\bibinfo {volume}
  {D76}},\ \bibinfo {pages} {054506} (\bibinfo {year} {2007})},\ \Eprint
  {http://arxiv.org/abs/0704.1111} {arXiv:0704.1111 [hep-th]} \BibitemShut
  {NoStop}%
\bibitem [{\citenamefont {Sexty}(2014)}]{Sexty}%
  \BibitemOpen
  \bibfield  {author} {\bibinfo {author} {\bibfnamefont {D.}~\bibnamefont
  {Sexty}},\ }\bibfield  {booktitle} {\emph {\bibinfo {booktitle}
  {{Proceedings, 24th International Conference on Ultra-Relativistic
  Nucleus-Nucleus Collisions (Quark Matter 2014)}}},\ }\href {\doibase
  10.1016/j.nuclphysa.2014.09.029} {\bibfield  {journal} {\bibinfo  {journal}
  {Nucl. Phys.}\ }\textbf {\bibinfo {volume} {A931}},\ \bibinfo {pages} {856}
  (\bibinfo {year} {2014})},\ \Eprint {http://arxiv.org/abs/1408.6767}
  {arXiv:1408.6767 [hep-lat]} \BibitemShut {NoStop}%
\bibitem [{\citenamefont {Bloch}(2011)}]{Bloch}%
  \BibitemOpen
  \bibfield  {author} {\bibinfo {author} {\bibfnamefont {J.}~\bibnamefont
  {Bloch}},\ }\href {\doibase 10.1103/PhysRevLett.107.132002} {\bibfield
  {journal} {\bibinfo  {journal} {Phys. Rev. Lett.}\ }\textbf {\bibinfo
  {volume} {107}},\ \bibinfo {pages} {132002} (\bibinfo {year} {2011})},\
  \Eprint {http://arxiv.org/abs/1103.3467} {arXiv:1103.3467 [hep-lat]}
  \BibitemShut {NoStop}%
\bibitem [{\citenamefont {Alford}\ \emph {et~al.}(1999)\citenamefont {Alford},
  \citenamefont {Kapustin},\ and\ \citenamefont {Wilczek}}]{Alford}%
  \BibitemOpen
  \bibfield  {author} {\bibinfo {author} {\bibfnamefont {M.~G.}\ \bibnamefont
  {Alford}}, \bibinfo {author} {\bibfnamefont {A.}~\bibnamefont {Kapustin}}, \
  and\ \bibinfo {author} {\bibfnamefont {F.}~\bibnamefont {Wilczek}},\ }\href
  {\doibase 10.1103/PhysRevD.59.054502} {\bibfield  {journal} {\bibinfo
  {journal} {Phys. Rev.}\ }\textbf {\bibinfo {volume} {D59}},\ \bibinfo {pages}
  {054502} (\bibinfo {year} {1999})},\ \Eprint
  {http://arxiv.org/abs/hep-lat/9807039} {arXiv:hep-lat/9807039 [hep-lat]}
  \BibitemShut {NoStop}%
\bibitem [{\citenamefont {de~Forcrand}(2009)}]{Forcrand}%
  \BibitemOpen
  \bibfield  {author} {\bibinfo {author} {\bibfnamefont {P.}~\bibnamefont
  {de~Forcrand}},\ }\bibfield  {booktitle} {\emph {\bibinfo {booktitle}
  {{Proceedings, 27th International Symposium on Lattice field theory (Lattice
  2009)}}},\ }\href@noop {} {\bibfield  {journal} {\bibinfo  {journal} {PoS}\
  }\textbf {\bibinfo {volume} {LAT2009}},\ \bibinfo {pages} {010} (\bibinfo
  {year} {2009})},\ \Eprint {http://arxiv.org/abs/1005.0539} {arXiv:1005.0539
  [hep-lat]} \BibitemShut {NoStop}%
\bibitem [{\citenamefont {D'Elia}\ and\ \citenamefont {Negro}(2012)}]{Elia}%
  \BibitemOpen
  \bibfield  {author} {\bibinfo {author} {\bibfnamefont {M.}~\bibnamefont
  {D'Elia}}\ and\ \bibinfo {author} {\bibfnamefont {F.}~\bibnamefont {Negro}},\
  }\href {\doibase 10.1103/PhysRevLett.109.072001} {\bibfield  {journal}
  {\bibinfo  {journal} {Phys. Rev. Lett.}\ }\textbf {\bibinfo {volume} {109}},\
  \bibinfo {pages} {072001} (\bibinfo {year} {2012})},\ \Eprint
  {http://arxiv.org/abs/1205.0538} {arXiv:1205.0538 [hep-lat]} \BibitemShut
  {NoStop}%
\bibitem [{\citenamefont {Alles}\ and\ \citenamefont {Papa}(2008)}]{Alles1}%
  \BibitemOpen
  \bibfield  {author} {\bibinfo {author} {\bibfnamefont {B.}~\bibnamefont
  {Alles}}\ and\ \bibinfo {author} {\bibfnamefont {A.}~\bibnamefont {Papa}},\
  }\href {\doibase 10.1103/PhysRevD.77.056008} {\bibfield  {journal} {\bibinfo
  {journal} {Phys. Rev.}\ }\textbf {\bibinfo {volume} {D77}},\ \bibinfo {pages}
  {056008} (\bibinfo {year} {2008})},\ \Eprint {http://arxiv.org/abs/0711.1496}
  {arXiv:0711.1496 [cond-mat.stat-mech]} \BibitemShut {NoStop}%
\bibitem [{\citenamefont {Alles}\ \emph {et~al.}(2014)\citenamefont {Alles},
  \citenamefont {Giordano},\ and\ \citenamefont {Papa}}]{Alles2}%
  \BibitemOpen
  \bibfield  {author} {\bibinfo {author} {\bibfnamefont {B.}~\bibnamefont
  {Alles}}, \bibinfo {author} {\bibfnamefont {M.}~\bibnamefont {Giordano}}, \
  and\ \bibinfo {author} {\bibfnamefont {A.}~\bibnamefont {Papa}},\ }\href
  {\doibase 10.1103/PhysRevB.90.184421} {\bibfield  {journal} {\bibinfo
  {journal} {Phys. Rev.}\ }\textbf {\bibinfo {volume} {B90}},\ \bibinfo {pages}
  {184421} (\bibinfo {year} {2014})},\ \Eprint {http://arxiv.org/abs/1409.1704}
  {arXiv:1409.1704 [hep-lat]} \BibitemShut {NoStop}%
\bibitem [{\citenamefont {Aoki}\ \emph {et~al.}(2008)\citenamefont {Aoki},
  \citenamefont {Horsley}, \citenamefont {Izubuchi}, \citenamefont {Nakamura},
  \citenamefont {Pleiter}, \citenamefont {Rakow}, \citenamefont {Schierholz},\
  and\ \citenamefont {Zanotti}}]{Horsley}%
  \BibitemOpen
  \bibfield  {author} {\bibinfo {author} {\bibfnamefont {S.}~\bibnamefont
  {Aoki}}, \bibinfo {author} {\bibfnamefont {R.}~\bibnamefont {Horsley}},
  \bibinfo {author} {\bibfnamefont {T.}~\bibnamefont {Izubuchi}}, \bibinfo
  {author} {\bibfnamefont {Y.}~\bibnamefont {Nakamura}}, \bibinfo {author}
  {\bibfnamefont {D.}~\bibnamefont {Pleiter}}, \bibinfo {author} {\bibfnamefont
  {P.~E.~L.}\ \bibnamefont {Rakow}}, \bibinfo {author} {\bibfnamefont
  {G.}~\bibnamefont {Schierholz}}, \ and\ \bibinfo {author} {\bibfnamefont
  {J.}~\bibnamefont {Zanotti}},\ }\href@noop {} {\  (\bibinfo {year} {2008})},\
  \Eprint {http://arxiv.org/abs/0808.1428} {arXiv:0808.1428 [hep-lat]}
  \BibitemShut {NoStop}%
\bibitem [{\citenamefont {Panagopoulos}\ and\ \citenamefont
  {Vicari}(2011)}]{Vicari}%
  \BibitemOpen
  \bibfield  {author} {\bibinfo {author} {\bibfnamefont {H.}~\bibnamefont
  {Panagopoulos}}\ and\ \bibinfo {author} {\bibfnamefont {E.}~\bibnamefont
  {Vicari}},\ }\href {\doibase 10.1007/JHEP11(2011)119} {\bibfield  {journal}
  {\bibinfo  {journal} {JHEP}\ }\textbf {\bibinfo {volume} {11}},\ \bibinfo
  {pages} {119} (\bibinfo {year} {2011})},\ \Eprint
  {http://arxiv.org/abs/1109.6815} {arXiv:1109.6815 [hep-lat]} \BibitemShut
  {NoStop}%
\bibitem [{\citenamefont {Azcoiti}\ \emph {et~al.}(2002)\citenamefont
  {Azcoiti}, \citenamefont {Di~Carlo}, \citenamefont {Galante},\ and\
  \citenamefont {Laliena}}]{Azcoiti}%
  \BibitemOpen
  \bibfield  {author} {\bibinfo {author} {\bibfnamefont {V.}~\bibnamefont
  {Azcoiti}}, \bibinfo {author} {\bibfnamefont {G.}~\bibnamefont {Di~Carlo}},
  \bibinfo {author} {\bibfnamefont {A.}~\bibnamefont {Galante}}, \ and\
  \bibinfo {author} {\bibfnamefont {V.}~\bibnamefont {Laliena}},\ }\href
  {\doibase 10.1103/PhysRevLett.89.141601} {\bibfield  {journal} {\bibinfo
  {journal} {Phys. Rev. Lett.}\ }\textbf {\bibinfo {volume} {89}},\ \bibinfo
  {pages} {141601} (\bibinfo {year} {2002})},\ \Eprint
  {http://arxiv.org/abs/hep-lat/0203017} {arXiv:hep-lat/0203017 [hep-lat]}
  \BibitemShut {NoStop}%
\bibitem [{\citenamefont {Vainshtein}\ \emph {et~al.}(1982)\citenamefont
  {Vainshtein}, \citenamefont {Zakharov}, \citenamefont {Novikov},\ and\
  \citenamefont {Shifman}}]{Instanton}%
  \BibitemOpen
  \bibfield  {author} {\bibinfo {author} {\bibfnamefont {A.~I.}\ \bibnamefont
  {Vainshtein}}, \bibinfo {author} {\bibfnamefont {V.~I.}\ \bibnamefont
  {Zakharov}}, \bibinfo {author} {\bibfnamefont {V.~A.}\ \bibnamefont
  {Novikov}}, \ and\ \bibinfo {author} {\bibfnamefont {M.~A.}\ \bibnamefont
  {Shifman}},\ }\href {\doibase 10.1070/PU1982v025n04ABEH004533} {\bibfield
  {journal} {\bibinfo  {journal} {Sov. Phys. Usp.}\ }\textbf {\bibinfo {volume}
  {25}},\ \bibinfo {pages} {195} (\bibinfo {year} {1982})},\ \bibinfo {note}
  {[Usp. Fiz. Nauk136,553(1982)]}\BibitemShut {NoStop}%
\bibitem [{\citenamefont {Coleman}(1988)}]{Coleman}%
  \BibitemOpen
  \bibfield  {author} {\bibinfo {author} {\bibfnamefont {S.}~\bibnamefont
  {Coleman}},\ }\href@noop {} {\emph {\bibinfo {title} {Aspects of symmetry:
  selected Erice lectures}}}\ (\bibinfo  {publisher} {Cambridge University
  Press},\ \bibinfo {year} {1988})\BibitemShut {NoStop}%
\bibitem [{\citenamefont {Olver}(1954)}]{Bessel}%
  \BibitemOpen
  \bibfield  {author} {\bibinfo {author} {\bibfnamefont {F.~W.~J.}\
  \bibnamefont {Olver}},\ }\href {\doibase 10.1098/rsta.1954.0021} {\bibfield
  {journal} {\bibinfo  {journal} {Phil. Trans. Roy. Soc. Lond.}\ }\textbf
  {\bibinfo {volume} {A247}},\ \bibinfo {pages} {328} (\bibinfo {year}
  {1954})}\BibitemShut {NoStop}%
\bibitem [{\citenamefont {Arfken}\ \emph {et~al.}(2011)\citenamefont {Arfken},
  \citenamefont {Weber},\ and\ \citenamefont {Harris}}]{Arfken}%
  \BibitemOpen
  \bibfield  {author} {\bibinfo {author} {\bibfnamefont {G.~B.}\ \bibnamefont
  {Arfken}}, \bibinfo {author} {\bibfnamefont {H.~J.}\ \bibnamefont {Weber}}, \
  and\ \bibinfo {author} {\bibfnamefont {F.~E.}\ \bibnamefont {Harris}},\
  }\href@noop {} {\emph {\bibinfo {title} {Mathematical methods for physicists:
  A comprehensive guide}}}\ (\bibinfo  {publisher} {Academic press},\ \bibinfo
  {year} {2011})\BibitemShut {NoStop}%
\end{thebibliography}%

\end{document}